\documentclass[%
reprint,
 amsmath,amssymb,
 aps,
]{revtex4-1}

\usepackage{graphicx}
\usepackage{dcolumn}
\usepackage{bm}
\usepackage[svgnames]{xcolor}
\usepackage{ulem}
\usepackage{xstring}
\usepackage{hyperref}
\usepackage{wasysym}

\allowdisplaybreaks

\newcommand{\rcite}[1]{\IfSubStr{#1}{,}{Refs.~}{Ref.~}\cite{#1}}

\hypersetup{
     colorlinks = true,
     linkcolor = red,
     citecolor = green,
     urlcolor = blue
     }

\begin{document}

\global\long\def\Appell{\,\text{F}_{1}^{\text{A}}}
\global\long\def\GaussF{\,_{2}\text{F}_{1}}

\preprint{APS/123-QED}

\title{Pulsar timing array signals induced by black hole binaries in relativistic eccentric orbits}

\author{Abhimanyu Susobhanan}
\email[Email: ]{s.abhimanyu@tifr.res.in}
 \affiliation{Department of Astronomy and Astrophysics, Tata Institute of Fundamental Research, Mumbai 400005, Maharashtra, India}

\author{Achamveedu Gopakumar}%
\affiliation{Department of Astronomy and Astrophysics, Tata Institute of Fundamental Research, Mumbai 400005, Maharashtra, India}%

\author{George Hobbs}
\affiliation{CSIRO Astronomy and Space Science, Australia Telescope National Facility, Box 76, Epping, NSW 1710, Australia}%

\author{Stephen R. Taylor}
\affiliation{Department of Physics \& Astronomy, Vanderbilt University, 2301 Vanderbilt Place, Nashville, TN 37235, USA}

\date{\today}

\begin{abstract}
Individual supermassive black hole binaries in non-circular orbits are possible nanohertz gravitational wave sources for the rapidly maturing Pulsar Timing Array  experiments. 
We develop an accurate and efficient approach to compute Pulsar Timing Array signals due to gravitational waves from inspiraling supermassive black hole binaries in relativistic eccentric orbits. 
Our approach employs a Keplerian-type parametric solution to model third post-Newtonian accurate precessing eccentric orbits while a novel semi-analytic prescription is provided to model the effects of quadrupolar order gravitational wave emission. 
These inputs lead to a semi-analytic prescription to model 
such signals, induced by 
non-spinning black hole binaries inspiralling along arbitrary eccentricity orbits.
Additionally, we provide a fully analytic prescription to model Pulsar Timing Array signals from black hole binaries inspiraling along moderately eccentric 
orbits, influenced by Boetzel {\it et al.} [Phys. Rev. D 96,044011(2017)]. 
These approaches are being incorporated into 
\texttt{Enterprise}  and  \texttt{TEMPO2} for searching the presence of such binaries in  Pulsar Timing Array datasets.
\end{abstract}


\maketitle


\section{Introduction}
\label{sec:intro}
Pulsar Timing Array (PTA) experiments are expected to inaugurate the field of nanohertz gravitational wave (GW) astronomy during the next decade \citep{Kelley2017}.
This will augment the ground-based GW astronomy, established by the LIGO Scientific Collaboration and the Virgo collaboration during the present decade, operating mainly in the hectohertz to kilohertz 
frequency regime  \citep{LIGO_Virgo_2018_Catalog}. 
A PTA experiment monitors  an ensemble of millisecond pulsars (MSPs) to search for correlated deviations to their pulse times of arrival (TOAs) to infer the presence of GWs \citep{FosterBacker1990}.
These efforts are sensitive to long-wavelength ($\sim$1 nHz -- 100 nHz) GWs, where the lower and upper limits of the frequency range are respectively set by the total span and cadence of the PTA observations \citep{Lommen2015}.
Therefore, PTAs are expected to detect GWs from supermassive black hole (SMBH) binaries with milliparsec orbital separations \citep{Detweiler1979}. 
At present, there exist three matured PTA efforts, namely the Parkes Pulsar Timing Array (PPTA) \citep{Hobbs2013_PPTA}, the European Pulsar Timing Array (EPTA) \citep{KramerChampion2013_EPTA}, and the North American Nanohertz Observatory for Gravitational Waves (NANOGrav) \citep{McLaughlin2013_nanograv,Brazier2019}.
Additionally, there are emerging PTA efforts from India, China, and South Africa \citep{Joshi2018_InPTA,Bailes2018,Lee2016}. 
The International Pulsar Timing Array (IPTA) consortium  combines  data and resources to  achieve more quickly the first detection of nanohertz GWs \citep{Hobbs2010_IPTA, Verbiest2016, Perera2019_IPTADR2}.

There are a number of promising astrophysical and cosmological GW sources in the nanohertz frequency window \citep{HobbsDai2017,Burke-Spolaor2019}.
We expect the first detected signal to be the ensemble of GWs from many SMBH binaries, producing a stochastic GW background (SGWB). 
This should be followed by the detection of bright individual SMBH binaries that resound above this background \cite{Rosado2015,Kelley2018}.
Stringent observational constraints are being placed on both types of PTA sources due to the absence of any firm detections in the  PTA datasets \citep{Arzoumanian2018_nanograv11_SGWB, Aggarwal2019_nanograv11_SMBHB, Babak2015_EPTA, Lentati2015_EPTA_GWB,Feng2019,Porayko2018}.
In the case of SMBH binaries in circular orbits, 
the present sky-averaged upper limit on GW strain is below $7.3 \times 10^{-15}$ at $8$ nHz \citep{Aggarwal2019_nanograv11_SMBHB}. 

Such constraints on SMBH binaries can be invoked to restrict their astrophysical formation and evolution scenarios \citep{Middleton2018,Chen2019,Chen2017,Taylor2019,Taylor2017}.
It will be desirable to extend the above bounds to eccentric binaries since SMBH binaries emitting nanohertz GWs can have non-negligible orbital eccentricities \citep{Burke-Spolaor2019}.
It was noted that SMBH binaries originating from gas-rich galaxy mergers may have non-negligible eccentricities even during their late inspiral phase \citep{ArmitageNatarajan2005,Cuadra2009}.
Additionally, realistic N-body simulations of massive galaxy mergers  result in SMBH binaries in  eccentric orbits due to stellar interactions \citep{Berentzen2009, Khan2012, Khan2013,RoedigSesana2012}. 
Therefore, it will be interesting to probe the presence of such binaries in the existing PTA datasets.
This demands general relativistic constructs that can be implemented  in the popular  pulsar timing software packages like  \texttt{TEMPO2} and  \texttt{Enterprise} \citep{Edwards2006_tempo2, Hobbs2006_tempo2, Ellis2017_enterprise}. 

In the  present paper, we develop an accurate and efficient prescription to obtain PTA signals induced by isolated SMBH binaries inspiraling along general relativistic eccentric orbits.  
Our approach employs the post-Newtonian (PN) approximation which allows us to model black holes (BHs) as point particles \citep{Blanchet2014}.
Recall that PN approximation provides general relativistic corrections to Newtonian dynamics in powers of $({v}/{c})^2 \sim {GM}/{c^2r}$, where  $v$, $M$, and $r$ are respectively the relative velocity, total mass, and relative separation of a BH binary.
We let BH binaries  move in 3PN-accurate precessing eccentric orbits with the help of generalized quasi-Keplerian parametrization \citep{Memmesheimer2004}, where the 3PN-accurate description incorporates $(\frac{v}{c})^6$-order general relativistic corrections to Newtonian motion.
Additionally, we incorporate the effects of GW emission at the dominant quadrupolar order with the help of a GW phasing formalism, detailed in \rcite{Damour2006,KonigsdorfferGopakumar2006}, while adapting recent results from \rcite{Moore2018}.
This allows us to model PTA signals due to non-spinning SMBH binaries inspiraling along 3PN-accurate eccentric orbits in a semi-analytic manner.
The numerical treatments are required only to solve the PN-accurate Kepler equation and to integrate the resulting fractional pulsar frequency shift induced by passing GWs.
These considerations ensure that the prescription is general relativistically accurate and computationally efficient.
It turns out that the PN description is quite appropriate to model such PTA signals as the SMBH binaries are expected to merge at orbital frequencies outside the PTA frequency window \citep{Burke-Spolaor2019}.
Additionally, we provide a fully-analytic prescription to compute PTA signals induced by  isolated SMBH binaries inspiraling along moderately eccentric orbits.
This result heavily depends on a fully analytic approach to compute temporally evolving GW polarization states for compact binaries moving in PN-accurate moderately eccentric orbits \citep{Boetzel2017}.
We note in passing  that the present effort extends and improves efforts to compute PTA signals due to GWs from compact binaries inspiraling along Newtonian accurate eccentric orbits \citep{Jenet2004, Taylor2016}. 

In what follows, we list below the salient features of the present paper.
\begin{itemize}
    \item 
    A brief description of our approach for computing quadrupolar-order PTA signals due to      inspiral GWs from non-spinning massive BH binaries in PN-accurate 
    precessing eccentric orbits while employing the PN-accurate Keplerian-type parametric solution and the GW phasing approach of Ref.~\cite{Damour2006}  is presented 
    in Sec.~\ref{sec:EccentricResiduals_expr} and \ref{sec:qKP}.
    \item
    An accurate and computationally efficient way to incorporate the effects of GW emission on the parametrized conservative PN-accurate orbital dynamics and its salient features are presented in Sec.~\ref{sec:orbital_evolution}. This subsection explains why we  require a one-time numerical solution of a differential equation to incorporate the effects of quadrupolar GW emission in our approach. Plots displaying PTA signals that arise from our semi-analytical approach and their various facets are 
    provided in Sec.~\ref{sec:illustrated_results}.
    The computational costs associated with our modeling of the PTA signals, induced by      GWs     from  massive BH binaries in PN-accurate arbitrary eccentricity orbits are provided in
    Sec.~\ref{sec:computation_cost}.
    
    \item  A fully analytic way of computing PTA signals for moderate eccentricities ($e\lesssim0.3$) is presented in Sec.~\ref{sec:analytic_residuals} where we employed crucial inputs 
    from Ref.~\cite{Boetzel2017}.
     This approach provides a powerful check on our detailed 
     semi-analytic prescription and this is demonstrated by comparing PTA signals computed using our semi-analytic and fully analytic methods in the low-eccentricity regime.
\end{itemize}


In brief,  we developed an accurate and efficient prescription to compute PTA signals induced by isolated SMBH binaries inspiraling along general relativistic eccentric orbits, employing for the first time an accurate semi-analytic solution to describe PN-accurate orbital evolution of BH binaries.
An implementation of the PTA signals derived in this work is available at \url{https://github.com/abhisrkckl/GWecc}.

\section{PTA signals from BH binaries in quasi-Keplerian eccentric orbits}
\label{sec:EccentricResiduals}
We begin by deriving  expressions for the dominant quadrupolar order $+/\times$ residuals in Sec.~\ref{sec:EccentricResiduals_expr}.
How we describe temporal evolution of various dynamical variables that appear in these expressions is described in Sec.~\ref{sec:qKP} and 
Sec.~\ref{sec:orbital_evolution}, which is followed by a  pictorial exploration of our main results in Sec.~\ref{sec:illustrated_results} and an exploration of the associated computational costs in  Sec.~\ref{sec:computation_cost}. 
The present paper explores the effects of far-zone GWs on the pulsar TOAs, and this is 
realistic as our GW sources are extra-galactic while the pulsars exist within our Galaxy.

\subsection{Timing residual expressions at the dominant quadrupolar order}
\label{sec:EccentricResiduals_expr}

When a GW signal passes across the line of sight between a pulsar and the observer  along a direction $\hat n$, it perturbs the underlying space-time metric. 
This induces temporally evolving changes in the measured pulsar rotational frequency $\nu$ \citep{BookFlanagan2011}
\begin{equation}
    \frac{\Delta \nu(t_{E})}{\nu}\equiv z_{\text{GW}}(t_{E}) = h(t_{E})-h(t_{P})\,,
\end{equation}
where $h$ stands for the dimensionless GW strain, $t_E$ and $t_P$ denote respectively the instances when a GW passes the solar system barycenter (SSB) and the pulsar, and the rotational frequency $\nu$ is measured in the SSB frame. 
These two time instances differ by the usual geometric delay such that 
\begin{align}
    t_P &= t_E - \frac{D_P}{c} \left( 1 + \hat n \cdot \hat p \right)\nonumber\\
    &= t_E-\frac{D_P}{c} (1-\cos \mu)
    \,,
\end{align}
where $D_P$ is the distance to the pulsar while $\hat p$ specifies its direction with respect to the SSB, and  $\mu$ provides the angle  between  $\hat n$ and  $\hat p$. 
Influenced by \rcite{Wahlquist1987}, the GW strain $h$ can be written in terms of the two GW polarization states $h_{+,\times}$ as 
\begin{equation}
    h(t)=
    \begin{bmatrix}F_{+} & F_{\times}\end{bmatrix}
    \begin{bmatrix}\cos2\psi & -\sin2\psi\\\sin2\psi & \cos2\psi\end{bmatrix}
    \begin{bmatrix}h_{+}(t)\\h_{\times}(t) \end{bmatrix}\,,
    \label{eq:h(t)_mat}
\end{equation}
where $F_{+,\times}$ are the antenna pattern functions that depend on the sky locations of the pulsar and the GW source, and $\psi$ is the polarization angle of the GW.
The explicit expressions for $F_{+,\times}$ involve angles that specify the directions $\hat n$ and  $\hat  p$ (namely, the right ascension (RA) and declination (DEC) of the GW source and the pulsar), and are available in \rcite{Lee2011}.


The temporally evolving GW-induced redshift causes differences between the expected and the observed TOAs of pulses.
This is given by   
\begin{align}
    R(t_E) &= \int_{0}^{t_E} z_{\text{GW}}(t')\,dt'  = s(t_E) - s(t_P)\,,
    \label{eq:R(t) int}
\end{align}
where $s(t)$ is given by
\begin{align}
    s(t) &= \int_0^t h(t') dt' = F_{+} s_{+}(t) + F_{\times} s_{\times}(t)\,,
    \label{eq:s(t)}
\end{align}
and we have defined 
\begin{equation}
  s_{+,\times}(t)=\int_0^t h_{+,\times}(t')\,dt'  \,.
  \label{eq:spx(t) int}
\end{equation}
This quantity $R(t_E)$ is usually referred to as the GW-induced (pre-fit) pulsar timing residual or the PTA signal, and is essentially prescribed by the values of $s_{+,\times}$ at the SSB and the pulsar positions.
It is customary to refer to  $s(t_E)$ and $s(t_P)$ as the Earth and pulsar terms, and $s_{+,\times}$ as the plus/cross residuals, respectively.

The leading quadrupolar order $h_{+,\times}$  expressions 
for a non-spinning eccentric binary, available in \rcite{Boetzel2017}, read 
\begin{widetext}
\begin{subequations}
\begin{align}
    h_{+}^{Q} &= \frac{GM\eta}{D_{L} \,c^2} x \frac{1}{(1-\chi)^{2}}\left(-2\left(c_{i}^{2}+1\right)\sqrt{1-e_{t}^{2}}\xi\sin(2\phi)+\left(c_{i}^{2}+1\right)\left(2e_{t}^{2}-\chi^{2}+\chi-2\right)\cos(2\phi)+s_{i}^{2}(1-\chi)\chi\right)\,,\\
    h_{\times}^{Q} &= \frac{GM\eta}{D_{L} \,c^2} x \frac{1}{(1-\chi)^{2}}2c_{i}\left(2\sqrt{1-e_t^{2}}\,\xi\cos(2\phi)+\left(2e_t^{2}-\chi^{2}+\chi-2\right)\sin(2\phi)\right)\,, 
\end{align}
\label{eq:h+x_N}
\end{subequations}
\end{widetext}
where $\phi$ denotes the angular coordinate in the orbital plane, called the orbital phase (see Eqs. \ref{eq:phi_1PN} and \ref{eq:phi_PN} below for the definition of $\phi$ for Newtonian and PN-accurate orbits) while the superscript $Q$ indicates the quadrupolar order contributions to $h_{+,\times}$.
The total mass, symmetric mass ratio and luminosity distance to the binary are represented by 
$M=m_1 + m_2$,
$\eta=\frac{m_1 m_2}{M^2}$,  and
$D_{L} $ respectively.
Further, we use shorthand notations to denote trigonometric functions of the orbital inclination $i$, namely
$c_i=\cos i$ and $s_i=\sin i$, while 
$\chi=e_t \cos u$ and $\xi=e_t \sin u$, where $u$ is the eccentric anomaly. 
The orbital eccentricity is specified by $e_t$ and it is associated with the PN-accurate Kepler equation \citep{Memmesheimer2004}. 
The dimensionless PN parameter $x = (G M n/c^3)^{2/3}$ employs the mean motion $n$ associated with the Kepler equation, which is related to the orbital period $P_b$ by $n=2\pi/P_b$.  
{
In addition, the polarization angle $\psi$ present in Eq.~(\ref{eq:h(t)_mat}) provides a measure of the longitude of the ascending node in the case of non-spinning binaries.
}


To obtain Eqs.~(\ref{eq:h+x_N}), we begin from the quadrupolar order $h_{+,\times}$ expressions that are valid for compact binaries in non-circular orbits \cite{Damour2006}
\begin{widetext}
\begin{subequations}
\begin{align}
h_{+}^{Q}(r,\phi,\dot{r},\dot{\phi})&=-\frac{GM\eta}{D_{L}c^{4}}\left[\left(1+c_{i}^{2}\right)\left[\left(\frac{GM}{r}+r^{2}\dot{\phi}^{2}-\dot{r}^{2}\right)\cos2\phi+2r\dot{r}\dot{\phi}\sin2\phi\right]+s_{i}^{2}\left(\frac{GM}{r}-r^{2}\dot{\phi}^{2}-\dot{r}^{2}\right)\right]\,,\\
h_{\times}^{Q}(r,\phi,\dot{r},\dot{\phi})&=-\frac{GM\eta}{D_{L}c^{4}}2c_{i}\left[\left(\frac{GM}{r}+r^{2}\dot{\phi}^{2}-\dot{r}^{2}\right)\sin2\phi-2r\dot{r}\dot{\phi}\cos2\phi\right]\,,
\end{align}
\label{eq:hxpQNC}
\end{subequations}
\end{widetext}
where $r$ and $ \phi$ provide the radial and angular coordinates that specify the position of the reduced mass $m_1 m_2/M$ around the total mass $M$ in the  center of mass frame of the binary, while $\dot r = d r/dt$ and $ \dot \phi = d \phi/dt$.
We employ the Keplerian parametric solution for eccentric orbits to provide parametric expressions for these dynamical variables. 
The classical Keplerian parametric solution, neatly summarized in \rcite{DamourDeruelle1985}, provides the following parametric expressions for $r$ and $\phi$:
\begin{subequations}
\begin{align}
r &= a \, \left ( 1 - e \, \cos u \right ) \,,\\
\phi - \phi_0 &= f \label{eq:phi_1PN}\,, 
\end{align}
\end{subequations}
where $a$ and $e$ specify respectively the orbital semi-major axis and 
the Newtonian orbital eccentricity such that $0\le e< 1$,
while $\phi_0$ is some initial orbital phase. The true anomaly
$f$ is related to the eccentric anomaly $u$ by the relation
\begin{equation}
    f=2\arctan\left[\left(\frac{1+e}{1-e}\right)^{1/2}\tan\frac{u}{2}\right]\,.
\end{equation}
This approach provides temporal evolution for $r$ and $\phi$ in a semi-analytic manner as  $u$ is related to the coordinate time $t$ by the transcendental Kepler equation \citep{DamourDeruelle1985}
\begin{equation}
l \equiv n(t - t_0) = u - e  \sin u \,,
\label{eq:KE_1PN}
\end{equation}
where $l$ is called the mean anomaly and $t_0$ denotes the epoch of periapsis passage. 

With the help of the such a parametric solution, it is fairly easy to obtain expressions for $r$,  $\dot r$, and $\dot \phi$ in terms of $u$, $e$ and $x$. 
This essentially leads to Eqs.~(\ref{eq:h+x_N}) from Eqs.~(\ref{eq:hxpQNC}) for $h_{+,\times}^{Q}$.
Note that we need an accurate and efficient method to tackle the above transcendental Eq.~(\ref{eq:KE_1PN}) to obtain the actual temporal evolution for the two polarization states. We note in passing that Eqs.~(\ref{eq:h+x_N}) and (\ref{eq:hxpQNC}) are also invoked to obtain inspiral templates for stellar mass compact binaries in eccentric binaries \cite{Tanay2016,Tiwari2019}.


In the next subsection, we summarize our approach to provide fully 3PN-accurate temporal evolution for 
our $h_{+,\times}^{Q}$ expressions.

\subsection{Accurate description for the evolution of non-spinning BH binaries inspiraling along precessing eccentric orbits}
\label{sec:qKP}

\begin{figure*}
    \centering
    \includegraphics[scale=0.56]{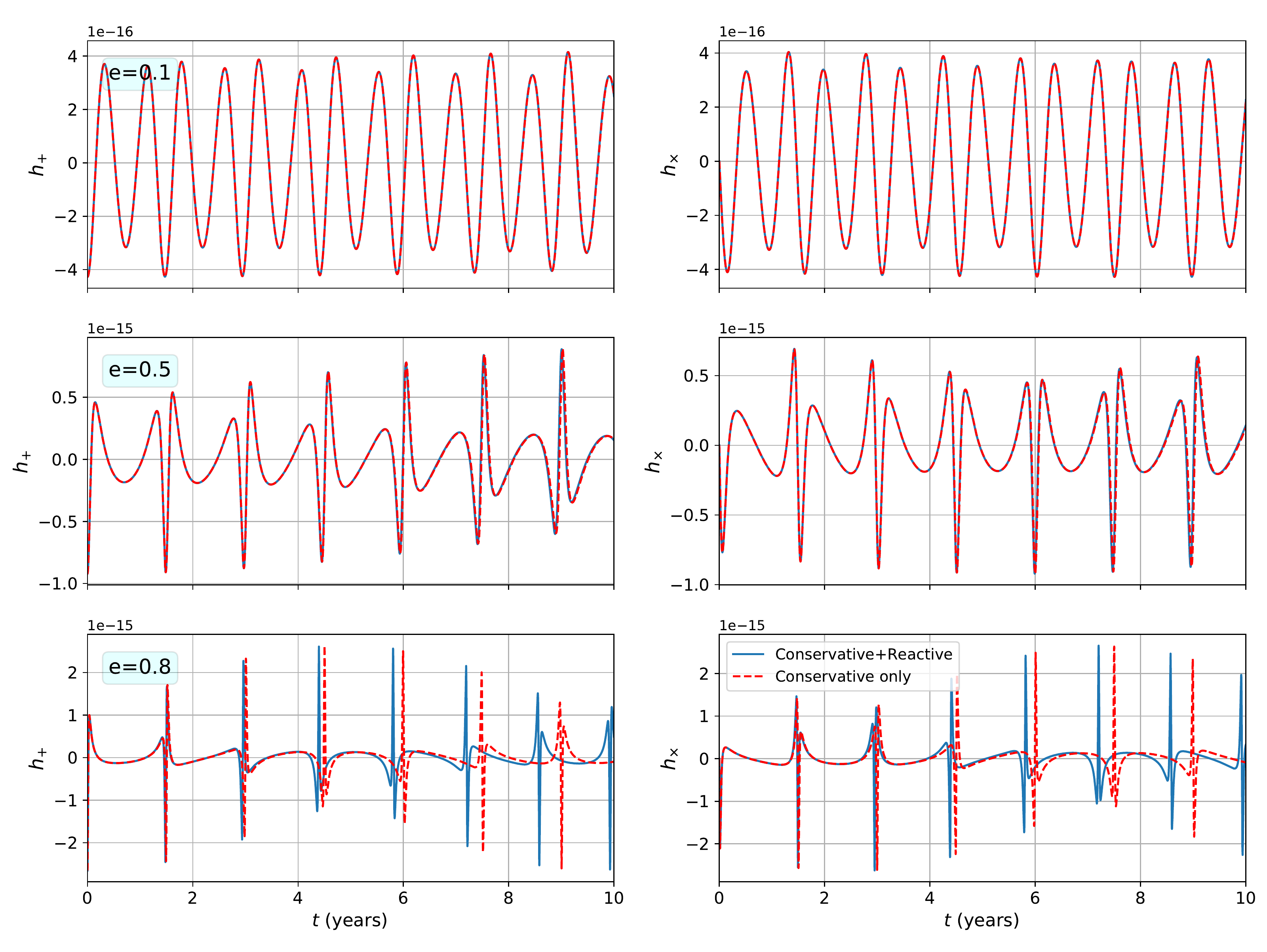}
    \caption{
Temporally evolving $h_{+,\times}^Q$, namely 
quadrupolar order GW polarization states, associated with massive BH binaries in 
3PN-accurate eccentric orbits.
The sold line plots incorporate the effects of GW emission that enter 
the binary BH dynamics at the 2.5PN order and we ignore the effects of gravitational radiation reaction in the dashed line plots.
We let $M=10^9 M_{\astrosun}$, $\eta=0.25$, $D_L=1$ Gpc,  $P_b=1.5$ years, $i=0$ and $\psi=0$ while selecting three  $e_t$ values (the listed $e_t$ values provide orbital eccentricities at $t_E=0$ epoch in all our figures).
The periapsis advance-induced amplitude modulations are clearly visible in the plots for moderately high to high eccentric binaries while GW emission-induced {\it chirping} is apparent in the $e=0.8$ plots.
}
\label{fig:hpx_comparison}
\end{figure*}

We begin by outlining our approach to describe the orbital evolution of non-spinning BH binaries inspiraling along 3PN-accurate quasi-Keplerian eccentric orbits. 
This prescription is crucial to specify how the angular variables ($\phi, u$) and the orbital elements ($n, e_t$) vary in time while computing  $R(t)$ as evident from Eqs.~(\ref{eq:h+x_N}).
First, we adapt the GW phasing formalism, detailed in \rcite{Damour2006,KonigsdorfferGopakumar2006}, for computing temporally evolving $h_{+,\times}(t)$.
This approach involves splitting the orbital dynamics of compact binaries into certain conservative and reactive parts.
In the PN terminology, the conservative dynamics usually provides PN corrections that are even powers of $(v/c)$, while  reactive dynamics involves odd powers of $(v/c)$ beginning with ${\cal O}((v/c)^5)$ contributions.
Such a split is justified as the reactive effects due to GW emission first enter the orbital dynamics only at the $(v/c)^5$ (2.5PN) order and act in timescales much longer than the orbital period when the binary is not close to its merger.
This split also allows us to employ PN-accurate Keplerian-type parametric solution for describing the 3PN-accurate conservative orbital dynamics, detailed in \rcite{Memmesheimer2004}.
Extending Eq.~(\ref{eq:phi_1PN}) to 3PN order, we write the 3PN-accurate orbital phase as \begin{align}
\phi- \phi_0 &= ( 1 + k)\,l + W (u(l), n,e_t)\,, 
\end{align}
where the angular variable $W(u)$ is $2\pi$ periodic in $u$, and $k$ represents the advance of periapsis per orbit \cite{Damour2006,KonigsdorfferGopakumar2006}.
We do not display here the explicit 3PN-accurate expressions for $k$ and $W(u)$ 
in terms of $n$, $e_t$, $M$, and $\eta$.
However, these expressions in the modified harmonic gauge are available as Eqs.~(11b) and (25a-25h) in \rcite{KonigsdorfferGopakumar2006}. 
Clearly, we need to specify how $u$ varies with time to obtain 3PN-accurate temporal orbital phase evolution.
The  following 3PN-accurate Kepler Equation, which extends Eq.~(\ref{eq:KE_1PN}), provides the required ingredient 
\begin{equation}
    l = u - e_t \sin u + \mathfrak{F}_t(u)\,,
    \label{eq:KE_PN}
\end{equation}
where the explicit 3PN-accurate expression for $\mathfrak{F}_t(u)$ in terms of $u$, $n$, $e_t$, $M$, and $\eta$ is given by Eq.~(27) in \rcite{KonigsdorfferGopakumar2006}.
It is helpful to solve 
the above  equation by invoking an improved version of Mikkola's method
to obtain 3PN-accurate temporal phase evolution \cite{Tanay2016}.
Recall that Mikkola's method provides most accurate and efficient method to
solve classical Kepler Equation and determine $u(l)$ \cite{Mikkola87}.
For the present effort, it is rather 
convenient to re-write the above expression for the orbital phase as
\begin{align}
 \phi &= l+\gamma + (1+k)(f-l) + \mathfrak{F}_{\phi}(u)\,,
 \label{eq:phi_PN} 
\end{align}
where $\gamma- \gamma_0 = k\,n\, (t- t_0)$ tracks the evolution of the periapsis, and the true anomaly $f$ is given by
\begin{equation}
    f=2\arctan\left[\left(\frac{1+e_{\phi}}{1-e_{\phi}}\right)^{1/2}\tan\frac{u}{2}\right]\,,
\end{equation}
where $e_\phi$ is some angular eccentricity such that $0\le e_\phi < 1$. 
The explicit 3PN-accurate expression for $e_\phi$ in terms of $e_t$, $n$, $M$ and $\eta$ is available in \rcite{Damour2006}.
{
We note that the angular variable 
 $\gamma$ is not identical to the argument of periapsis $\omega$, usually defined 
 for Keplerian  orbits as $\phi - \omega = f$.
 This angular variable is termed as the  angle of periapsis
 and evolves as $ \gamma - \gamma_0 = k\,n\, ( t- t_0)$ for conservative PN orbits.
 }
Further, the definition of the mean anomaly $l$,
namely $l = n(t-t_0)$, ensures that both $l$ and $\gamma$ are linear-in-time varying angular variables.
Note that the use of Eqs.~(\ref{eq:KE_PN}) and (\ref{eq:phi_PN}) in our expressions for $h_{+,\times}^{Q}$, given by Eqs.~(\ref{eq:h+x_N}),  leads to an essentially analytic way for modeling temporally evolving quadrupolar GW polarization states.
The resulting waveforms are displayed as the dashed line plots in Fig.~\ref{fig:hpx_comparison} and we clearly see the periapsis advance-induced amplitude modulations in moderate to high eccentricity plots.
It is important to note that these dashed line plots provide $h_{+,\times}^{Q}$ associated with compact binaries moving in {\em conservative} 3PN-accurate precessing eccentric orbits.

Clearly, we need a prescription to include the effects of GW emission to model $h_{+,\times}(t)$ for compact binaries inspiraling along PN-accurate eccentric orbits.
This is pursued by adapting the GW phasing formalism of \rcite{Damour2006,KonigsdorfferGopakumar2006}.
This formalism demonstrated that GW emission forces $n$ and $e_t$ to change with time and it is possible to split their temporal evolution into two parts \cite{Damour2006}.
The first part leads to the secular or  orbital-averaged evolution equations for $n$ and $e_t$ which ensure that both $n$ and $e_t$ can change substantially over the gravitational radiation reaction timescale.
The second part essentially provides periodic variations to $n$ and $e_t$ in the orbital timescale, which remain tiny during the early inspiral phase of compact binary evolution \cite{Damour2006}. 
Therefore, we ignore such periodic variations to $n$ and $e_t$ for the present investigation as our focus is indeed on the early part of the BH binary inspiral. 
The secular evolution of $n$ and $e_t$ ensures that $l$ and $\gamma$ no longer follow linear-in-time variations as noted earlier. 
With the inclusion of gravitation radiation reaction effects, the explicit temporal evolution for $l$ and $\gamma$ becomes 
\begin{subequations}
\begin{align}
l-l_0 &= \int_{t_0}^t n(t')\, dt' \,,\\    
\gamma-\gamma_0 &= \int_{t_0}^t k(t')n(t')\, dt' \,,
\end{align}
\end{subequations}
where we have ignored orbital timescale variations in these angular variables \cite{Damour2006}.
These considerations imply that the GW phasing formalism provides a set of coupled differential equations 
for $n$, $e_t$, $\gamma$, and $l$. 
The resulting set of {\it four} coupled ordinary differential equations (ODEs) that incorporate secular effects of quadrupolar order GW emission read 
\cite{Damour2006},
\begin{subequations}
\begin{align}
  \frac{dn}{dt} &= \frac{1}{5}\left(\frac{GM_{\text{ch}}n}{c^{3}}\right)^{\frac{5}{3}}n^{2}\frac{\left(96+292e_{t}^{2}+37e_{t}^{4}\right)}{\left(1-e_{t}^{2}\right)^{7/2}}\,,\label{eq:ode_n} \\
  \frac{d e_t}{dt} &=  \frac{-1}{15}\left(\frac{GM_{\text{ch}}n}{c^3}\right)^{\frac{5}{3}} n^{}e_t\frac{\left(304+121e_t^{2}\right)}{\left(1-e_t^{2}\right)^{5/2}}\,,\label{eq:ode_e}\\
  \frac{d \gamma}{dt} &= k\,n \,, \label{eq:ode_gamma}\\
  \frac{d l}{dt} &= n \label{eq:ode_l}\,,
\end{align}
\label{eq:ode_system} 
\end{subequations}  
where  $M_{\text{ch}}=\eta^{3/5} M$ is the chirp mass of the binary.
Note that we are required to solve the above set of four differential equations along with 3PN-accurate expressions for $u$ and $\phi$, given by Eqs.~(\ref{eq:KE_PN}) and (\ref{eq:phi_PN}) to 
describe the orbital phase evolution of compact binaries inspiraling along 3PN-accurate eccentric orbits.
In the next subsection, we develop a method to tackle these coupled differential equations in an essentially semi-analytic way.

\subsection{Semi-analytic description for $n(t)$, $e_t(t)$, $\gamma(t)$, and $l(t)$}
\label{sec:orbital_evolution}

\begin{figure}
    \centering
    \includegraphics[scale=0.47]{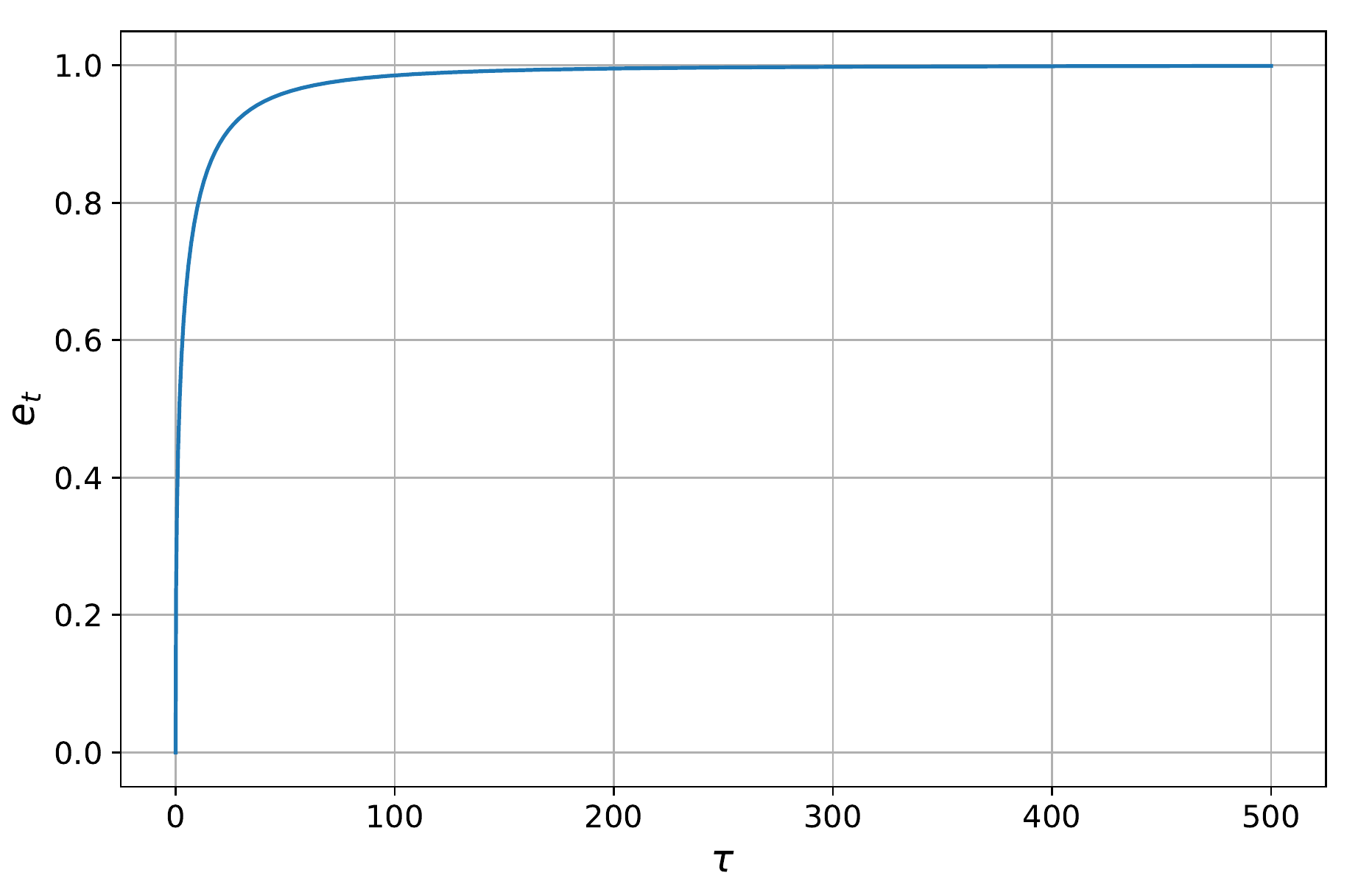}
    \caption{Numerical solution of Eq. (\ref{eq:de/dtau}) that provides $e_t(\tau)$ . 
    Note that the dimensionless temporal variable $\tau$ is defined as $\tau \equiv \tau_0 - \kappa (t-t_0)$ and this is why a compact binary evolves from right to left along the $\tau$ axis. 
    Let us emphasize that 
 this curve defines the orbital eccentricity evolution for all compact binaries and the origin provides 
    certain Newtonian merger epoch.
    }
    \label{fig:e(tau)}
\end{figure}

We begin by describing our computationally efficient way to obtain $n(t)$ and $e_t(t)$, influenced by \rcite{Damour2006,Moore2018}.
Our approach involves deriving certain analytic expressions for $n(e_t)$ and $t(e_t)$ and appropriately treating them numerically to  obtain an accurate and efficient way to track the temporal evolution in $n(t)$ and $e_t(t)$.
To obtain an analytic expression for $n(e_t)$, we divide  Eq.~(\ref{eq:ode_n}) by Eq.~(\ref{eq:ode_e}), and this leads to 
\begin{equation}
\frac{dn}{de_t}=-3\frac{n}{e_t}\frac{1}{\left(1-e_t^{2}\right)}\frac{\left(96+292e_t^{2}+37e_t^{4}\right)}{\left(304+121e_t^{2}\right)}\,.
\label{eq:dn_de}
\end{equation}
It is easy to integrate the above equation to obtain 
\begin{equation}
n(e_t)=n_{0}\left(\frac{e_{t0}}{e_t}\right)^{\frac{18}{19}}\left(\frac{1-e_t^{2}}{1-e_{t0}^{2}}\right)^{\frac{3}{2}}\left(\frac{304+121e_{t0}^{2}}{304+121e_t^{2}}\right)^{\frac{1305}{2299}}\,,
\label{eq:n(e)}
\end{equation}
where $n_0$ and $e_{t0}$ are the values of $n$ and $e_t$ at some initial epoch $t=t_0$ \cite{Damour2006}.
Unfortunately, it is not easy to obtain such a compact expression for $e_t(t)$. 
To obtain an equation that can be analytically tackled, we substitute the above equation for $n(e_t)$ in Eq.~(\ref{eq:ode_e}). 
The resulting equation may be written as
\begin{subequations}
\begin{equation}
\frac{de_t}{dt} =- \kappa\, \frac{\left(1-e_t^{2}\right)^{3/2}}{e_t^{29/19}\left(121e_t^{2}+304\right)^{1181/2299}}\,,
\label{eq:de/dt_one}
\end{equation}
where
\begin{equation}
\kappa = 
\frac{1}{15}\left(\frac{GM_{\text{ch}}n_0}{c^3}\right)^{\frac{5}{3}}
\frac{ n_0^{} e_{t0}^{\frac{48}{19}} \left(121 e_{t0}^2+304\right)^{\frac{3480}{2299}}}{\left(1-e_{t0}^2\right)^4}\,. \label{eq:P_coeff}    
\end{equation}
\end{subequations}
Note that the coefficient $\kappa$ is only a function of certain intrinsic binary BH parameters like the chirp mass, initial values of the mean motion and orbital eccentricity.
Further, it is not difficult to infer that $\kappa$ has the dimensions of frequency and  is non-zero for eccentric binaries.
These considerations influenced us to introduce a dimensionless temporal parameter $\tau$ such that $\tau = \tau_0 - \kappa \left( t- t_0 \right)$, and 
Eq.~(\ref{eq:de/dt_one}) in terms of $\tau$ becomes 
\begin{equation}
\frac{de_t}{d\tau}=\frac{\left(1-e_t^{2}\right)^{3/2}}{e_t^{29/19}\left(121e_t^{2}+304\right)^{1181/2299}}\,,
\label{eq:de/dtau}
\end{equation}
and we will clarify the significance  of the constant $\tau_0$ later.
Interestingly, this equation does not contain any intrinsic (and constant) binary BH parameters.
In other words, the above equation is valid for all eccentric compact binaries while restricting the GW emission effects to the leading quadrupolar order. 
It turns out that it is possible to obtain an analytical solution for Eq.~(\ref{eq:de/dtau}), as noted in \rcite{Moore2018},
and it reads 
\begin{equation}
\tau(e)=\frac{e^{\frac{48}{19}}}{768}\,\text{F}_{1}^{\text{A}}\left(\frac{24}{19};\frac{-1181}{2299},\frac{3}{2};\frac{43}{19};\frac{-121e^{2}}{304},e^{2}\right)\,,
 \label{eq:tau(e)_anl}
\end{equation}
where $\text{F}_{1}^{\text{A}}$ represents Appell's 2D hypergeometric function \cite{Colavecchia2001}, and we have chosen the initial condition $\tau(0)=0$ so that the constant of integration vanishes. 
It is indeed computationally very expensive to invert the above expression to get $e_t(\tau)$, mainly due to the difficulty in computing $\text{F}_{1}^{\text{A}}$ numerically.
Therefore, we pre-compute $e_t(\tau)$ at a sufficiently dense set of points and interpolate between those points to get $e_t(\tau)$ for arbitrary values of $\tau$.
Such a {\it look-up table} of $e_t(\tau)$ may be obtained either by numerically solving  Eq.~(\ref{eq:de/dtau}) or by inverting Eq.~(\ref{eq:tau(e)_anl}).
The resulting $e_t(\tau)$ plot is displayed in Fig.~\ref{fig:e(tau)} and it is important to note that GW emission forces $e_t$ to advance from right to left in our $e_t(\tau)$  plot.
This is essentially due to the way $\tau$ is related to the coordinate time $t$,
namely $\tau = \tau_0 - \kappa \left ( t- t_0 \right )$.
We have verified that our $e_t(\tau)$ evolution is consistent with Eq.~(51) of \rcite{Moore2018}.

We note here that the frequency $n\rightarrow \infty$ as $e_t\rightarrow 0$  as evident from Eq. (\ref{eq:n(e)}) and it influenced us to define certain merger time \textit{in our 2.5PN approximation} as the instant when $e_t\rightarrow 0$. 
We are now in a position to explain the meaning of $\tau_0$ and for this purpose, we define certain dimensionless {merger time} by invoking the initial condition $\tau(0)=0$.
This allows us to specify the above undetermined constant as $\tau_{0}=\tau(e_{t0})$, where $\tau(e_t)$ is given by Eq.~(\ref{eq:tau(e)_anl}).
We identify $\tau_0$ as certain dimensionless {merger time} because it is possible to compute certain `Newtonian' merger time for compact binaries with its help.  
The relevant expression for such a merger time is given by 
\begin{equation}
 t_{\text{merg}}^{\text{2.5PN}} = \frac{\tau_0}{\kappa} \,,  
\end{equation}
and we have verified that this expression, in the small eccentricity limit, is indeed consistent with Eq.~(50) of \rcite{Krolak1995}. 
Recall that \rcite{Krolak1995} computed the `Newtonian merger time' for compact binaries that incorporates the leading order 
eccentricity contributions as
{\small
\begin{equation}
\lim_{e_0 \rightarrow 0} t_{\text{merg}}^{\text{2.5PN}} 
\sim 
\frac{5}{n_{0}}\left(\frac{GM_{\text{ch}}n_0}{c^3}\right)^{-\frac{5}{3}}
\left( \frac{1}{256} - \frac{157e_0^2}{11008} \right)\,.
\label{eq:merger_time_ecc}
\end{equation}}
Additionally, we have computed an equivalent expression for such a merger time in  Appendix~\ref{sec:circular_orbits} while clarifying our way to treat the $\kappa \rightarrow 0$ scenario.

Note that as a binary BH approaches the $\tau =0$ epoch, its orbital dynamics becomes more relativistic and this eventually leads to the breakdown of the present quadrupolar (or 2.5PN) order description of the binary BH reactive dynamics.
Therefore, our prescription should only be used for an observational duration $t-t_0 $ which is substantially smaller than $t_{\text{merge}}^{\text{2.5PN}}$.
It turns out that our fully 3PN-accurate orbital description that incorporates the effects of quadrupolar order GW emission is quite appropriate while dealing with the expected isolated SMBH binary PTA sources. 
 
We now turn our attention to the evolution equations for $\gamma$ and $l$, given by Eqs. (\ref{eq:ode_gamma}) and (\ref{eq:ode_l}).
The plan is to express both $n$ and $k\,n$ in terms of $e_t$, $n_0$ and $e_0$ with the help of our $n(e_t)$ expression. 
Further, we employ our $\tau$ variable rather than its coordinate time ($t$) counterpart. This leads to 
\begin{subequations}
\begin{align}
   \frac{dl}{d\tau} &= -\alpha\frac{\left(1-e_t^{2}\right)^{\frac{3}{2}}}{e_t^{\frac{18}{19}}\left(304+121e_t^{2}\right)^{\frac{1305}{2299}}}\,, \label{eq:dl_dtau} \\
   \frac{d\gamma}{d\tau}&=-\beta\frac{\left(1-e_t^{2}\right)^{\frac{3}{2}}}{e_t^{\frac{30}{19}}\left(304+121e_t^{2}\right)^{\frac{2175}{2299}}}\,,\label{eq:dgamma_dtau}
\end{align}
\end{subequations}
where the dimensionless coefficients $\alpha$ and $\beta$ are given by 
\begin{subequations}
\begin{align}
    \alpha&=\left(\frac{GM_{\text{ch}}n_{0}}{c^{3}}\right)^{-\frac{5}{3}}\frac{15\left(1-e_{t0}^{2}\right){}^{5/2}}{e_{t0}^{\frac{30}{19}}\left(121e_{t0}^{2}+304\right)^{\frac{2175}{2299}}}\,,\\
    \beta&=\left(\frac{GM_{\text{ch}}n_{0}}{c^{3}}\right)^{-\frac{5}{3}}\left(\frac{GMn_{0}}{c^{3}}\right)^{\frac{2}{3}}\frac{45\left(1-e_{t0}^{2}\right){}^{3/2}}{e_{t0}^{\frac{18}{19}}\left(121e_{t0}^{2}+304\right)^{\frac{1305}{2299}}}\,.
\end{align}
\end{subequations}
It should be noted 
that we have only used the dominant order contributions to $k$, namely $k= 3\, x/(1-e_t^2)$, while obtaining the above equation for $d\gamma/dt$. Its 3PN 
extension is provided in Appendix \ref{sec:gamma_PN}.

 The next step is to obtain differential equations for $l$ and $\gamma$ that are independent of binary BH intrinsic (and constant) parameters.
To this end, we  define two  scaled and shifted variables $\bar{l}=L_{0}-l/\alpha $ and $\bar{\gamma}=\Gamma_{0}-\gamma/\beta$.
Invoking Eqs.~(\ref{eq:dl_dtau}) and (\ref{eq:dgamma_dtau}), it is fairly straightforward to obtain the following differential equations for $\bar l$ and $\bar \gamma$ 
\begin{subequations}
\begin{align}
    \frac{d\bar{l}}{d\tau}&=\frac{\left(1-e_t^{2}\right)^{\frac{3}{2}}}{e_t^{\frac{18}{19}}\left(304+121e_t^{2}\right)^{\frac{1305}{2299}}}\,, \label{eq:dlbar/dtau}\\
    \frac{d\bar{\gamma}}{d\tau}&=\frac{\left(1-e_t^{2}\right)^{\frac{3}{2}}}{e_t^{\frac{30}{19}}\left(304+121e_t^{2}\right)^{\frac{2175}{2299}}}\,, \label{eq:dgammabar/dtau}
\end{align}
\end{subequations}
with the following initial conditions ${l}(\tau_0)=l_0$ and ${\gamma}(\tau_0)=\gamma_0$.
 These initial conditions imply that the shifts $L_0$ and $\Gamma_0$ are given by
\begin{subequations}
\begin{align}
    L_{0}&=\bar{l}(\tau_{0}) + \frac{l_{0}}{\alpha} \,, \label{eq:L0} \\
    \Gamma_{0}&=\bar{\gamma}(\tau_{0})+\frac{\gamma_{0}}{\beta} \label{eq:Gamma0}\,.
\end{align}
\end{subequations}
The structure of the above two differential equations support analytic solutions if we compute $ d \bar l/ de_t$ and $ d \bar \gamma/ de_t$  versions of Eqs.~(\ref{eq:dlbar/dtau}) and (\ref{eq:dgammabar/dtau}) with the help of Eq. (\ref{eq:de/dtau}) for $d e_t/d\tau$. 
This results in 
\begin{subequations}
\begin{align}
    \frac{d\bar{l}}{de_t}&=\frac{e_t^{11/19}}{\left(121e_t^{2}+304\right)^{124/2299}}\,,\\
    \frac{d\bar{\gamma}}{de_t}&=\frac{e_t^{-1/19}}{\left(121e_t^{2}+304\right)^{994/2299}}\,.
\end{align}
\end{subequations}
The fact that the RHS of these equations depend only on $e_t$ allows us to obtain the following expressions for $\bar l$ and $\bar\gamma$ 
\begin{subequations}
\begin{align}
\bar{l}(e_t)&=\frac{19^{\frac{2175}{2299}}}{30\times2^{\frac{496}{2299}}}e_t^{\frac{30}{19}}\,_{2}\text{F}_{1}\left(\frac{124}{2299},\frac{15}{19};\frac{34}{19};\frac{-121e_t^{2}}{304}\right), \label{eq:lbar(et)}\\
\bar{\gamma}(e_t)&=\frac{19^{\frac{1305}{2299}}}{36\times2^{\frac{1677}{2299}}}e_t^{\frac{18}{19}}\,_{2}\text{F}_{1}\left(\frac{994}{2299},\frac{9}{19};\frac{28}{19};\frac{-121e_t^{2}}{304}\right) \label{eq:gammabar(et)},
\end{align}
\end{subequations}
where ${}_{2}\text{F}_{1}$ is the Gaussian hypergeometric function, and we have verified that the above expression for $\bar l$ is consistent with Eq.~(52) of \rcite{Moore2018}.
In Figure \ref{fig:lbar(tau),gammabar(tau)}, we plot these variables against $e_t$ and find the expected sharp rise in $\bar l$ for higher orbital eccentricities.
It is important to note that these plots are independent of the intrinsic (and constant) binary BH parameters like the total mass, mass ratio, and initial orbital eccentricity and period.

\begin{figure}
    \centering
    \includegraphics[scale=0.47]{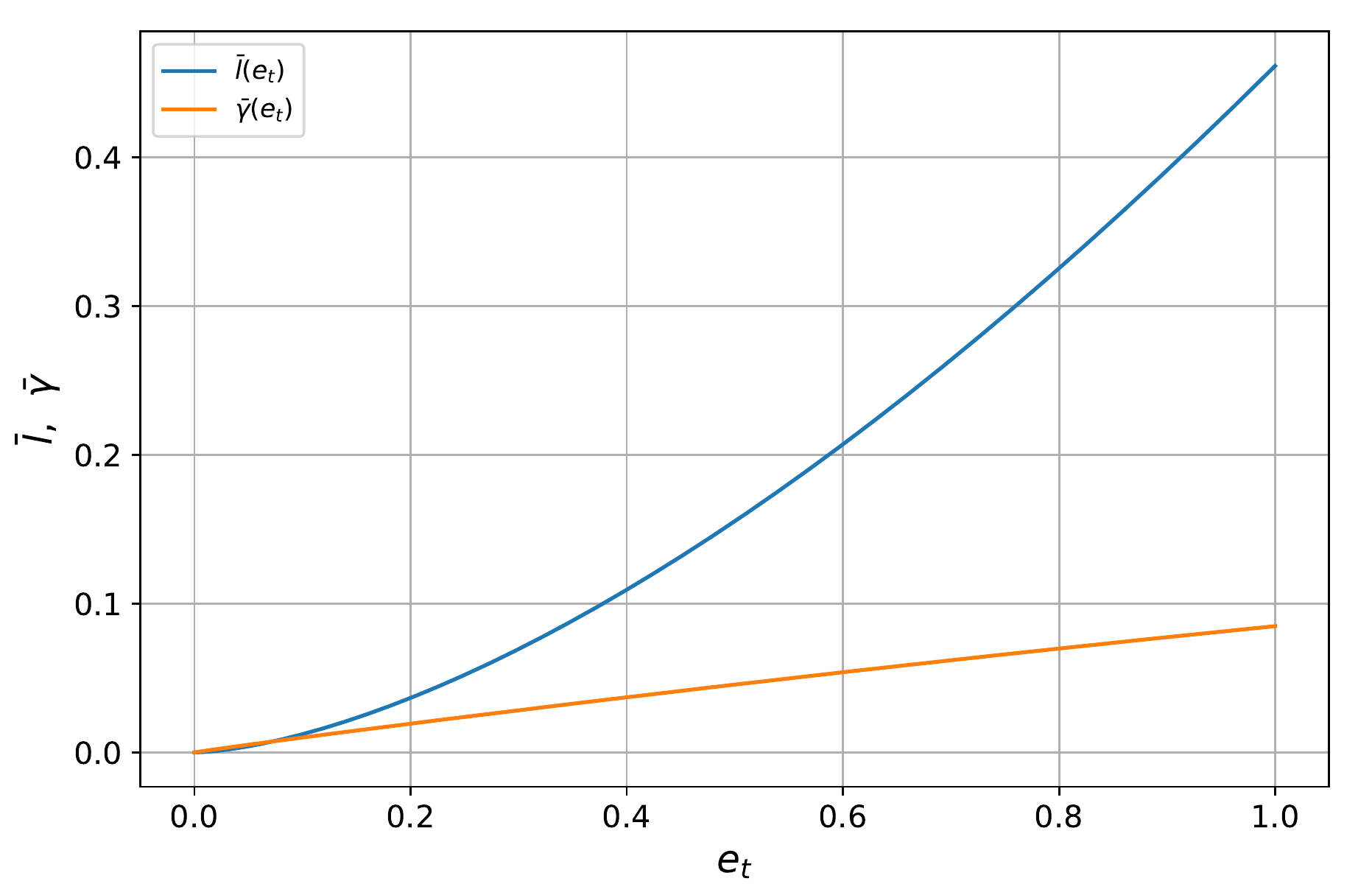}
    \caption{The non-linear variation of our two scaled angular variables $\bar{l}$ and $\bar{\gamma}$ as function of $e_t$. The use of these variables ensures that such variations are system independent at the quadrupolar order GW emission. }
    \label{fig:lbar(tau),gammabar(tau)}
\end{figure}

To obtain the actual temporal evolution for the above set of variables, namely $n$, $e_t$, $l$ and $\gamma$, we proceed as follows. 
First, we compute a look-up table for $e_t(\tau)$ by solving the the differential equation for $d e_t/d \tau$  as described earlier.
We emphasize here that this is a one-time computation since the differential equation (\ref{eq:de/dtau}) does not contain any system-dependent parameters, which implies that the look-up table, once computed, may be saved and re-used for later computations.
(Details of this computation are given in subsection \ref{sec:e(tau)_computation}.)
Thereafter, we determine $n(\tau)$, $l(\tau)$, and $\gamma (\tau)$ with the help of Eqs.~(\ref{eq:n(e)}), (\ref{eq:lbar(et)}), and (\ref{eq:gammabar(et)}) that involve hypergeometric functions.
Using the explicit expressions for $\tau_0$, $\kappa$, $\alpha$, and $\beta$ and specific relations that connect $\tau$ to $t$, $\bar l $ to $l$, and $\bar \gamma$ to $\gamma$, it is straightforward to obtain binary BH system-dependent temporal evolution for $n$, $e_t$, $l$, and $\gamma$ in terms of the regular coordinate time $t$. 
Let us emphasize  that these variable changes are easy to implement as they essentially involve analytic expressions. 
To ascertain the accuracy of this procedure, we compared $n(t)$, $e_t(t)$, $\gamma(t)$ and $l(t)$ computed using this method to results obtained by numerically solving the system of ODEs (\ref{eq:ode_system}) for different initial conditions, masses and mass ratios. We find that the results agree up to the numerical precision of the ODE solver as expected.

The variables $u$ and $\phi$ which appear in the waveform (\ref{eq:h+x_N}) may be computed using Eqs. (\ref{eq:KE_PN}-\ref{eq:phi_PN}). 
Finally, the PTA signal $R(t)$ can be computed by numerically integrating the waveform as given by Eqs. (\ref{eq:h(t)_mat}-\ref{eq:spx(t) int}).
We are forced to perform this integral numerically owing to the fact that the waveform (\ref{eq:h+x_N}) is a function of $u$ and $\phi$ which are not simple functions of the coordinate time.

\subsubsection{Computation of $e_t(\tau)$}
\label{sec:e(tau)_computation}
Clearly, an accurate and efficient prescription to 
obtain $e_t(\tau)$ is crucial for describing the temporal 
evolution of ($n,e_t,\gamma, l$) in terms of the 
coordinate time $t$.
The fact that an explicit expression is available only for $\tau(e_t)$ and not for $e_t(\tau)$ forced 
us to obtain $e_t(\tau)$ either by numerically integrating 
Eq.~(\ref{eq:de/dtau}) or by numerically inverting the analytic expression for $\tau(e_t)$ given by Eq.~(\ref{eq:tau(e)_anl}).
However, we pursued the relatively computationally inexpensive approach of computing a look-up table for $e_t(\tau)$ at a sufficiently dense sample of $\tau$ values for one time.
Thereafter, we obtain values of $e_t(\tau)$ at arbitrary $\tau$  values by  interpolating between the pre-computed values and this is heavily influenced by the universal nature of Eq.~(\ref{eq:de/dtau}).
In practice, we solve Eq.~(\ref{eq:de/dtau}) using an adaptive ODE solver, which adjusts the step size to ensure 
an optimal accuracy of the solution while constructing the look-up table.
This is important as 
the curvature of the function $e_t(\tau)$ is highly variable, as evident from Fig.~(\ref{fig:e(tau)}). Therefore,
 the look-up table must be computed at a non-uniform sample of points such that the regions of high curvature 
are sampled at sufficiently high density for ensuring high accuracy.

This approach poses 
 a new challenge since our  $de_t/d\tau$ equation diverges at $e_t=0$ as evident from Eq.~(\ref{eq:de/dtau}). 
This implies that the numerical integration cannot start  with the expected initial condition, namely $e_t(0)=0$. 
We avoid this issue 
by starting the numerical integration at a small non-zero value of $\tau$, say certain $\tau_{\min}$. 
To compute such an initial condition $e_t(\tau_{\min})$, we explore the asymptotic $(\tau,e_t)\rightarrow (0,0)$ behaviour of Eq.~(\ref{eq:de/dtau}). In this limit, Eq.~(\ref{eq:de/dtau}) becomes
\begin{align}
    \lim_{\tau,e_t \rightarrow 0} \frac{de_t}{d\tau}&\sim \frac{1}{304^{\frac{1181}{2299}}e_t^{\frac{29}{19}}} \,,
\end{align}
where we have expanded the R.H.S. of Eq.~(\ref{eq:de/dtau})  to the leading order contributions in $e_t$.
This equation can be integrated to obtain
\begin{align}
\lim_{\tau\rightarrow0}e_t(\tau)&\sim\frac{2^{559/726}3^{19/48}}{19^{145/242}}\tau^{19/48}\,.
\end{align}
Therefore, the new initial condition becomes 
\begin{equation}
    e_{t,\min} = \frac{2^{559/726}3^{19/48}}{19^{145/242}}\tau_{\min}^{19/48}\,,
\end{equation}
for some sufficiently small $\tau_{\min}$. 
The look-up table for $\tau(e)$ can now be computed by integrating Eq.~(\ref{eq:de/dtau}) from $\tau_{\min}$ to some $\tau_{\max}$ such that it covers all eccentricity values of interest.

It is also possible to provide an estimate for $\tau_{\max}$
where we can stop the numerical integration.
Using the fact that $\lim_{\tau\rightarrow\infty}e_t=1$, we  write Eq.~(\ref{eq:de/dtau}) in the $\tau\rightarrow\infty$ limit as
\begin{align}
 \lim_{\tau\rightarrow\infty}   \frac{d\epsilon}{d\tau}&\sim \frac{2\sqrt{2}\,\epsilon^{3/2}}{5\ 5^{63/2299}17^{1181/2299}}\,,
\end{align}
where we have substituted $\epsilon=1-e_t$ in Eq.~(\ref{eq:de/dtau}) and expanded the R.H.S. of the resulting equation to the leading order  in $\epsilon$.
This equation can be solved fairly easily to obtain
\begin{align}
   \lim_{\tau\rightarrow\infty} e_t&\sim 1-\frac{4}{(a\tau+b)^{2}}\,,
\end{align}
where we have defined the coefficient
\begin{align}
    a&=\frac{2\sqrt{2}}{5\times 5^{63/2299}17^{1181/2299}}\,.
\end{align}
In contrast, 
the coefficient $b$ may be computed by imposing the initial condition $e_t(\tau_{\max})=e_{t,\max}$
to be
\begin{align}
    b&=\frac{2}{\sqrt{1-e_{t,\max}}}-a\tau_{\max}\,.
\end{align}
In our approach, we provide these limits to obtain 
an accurate and efficient prescription to evaluate $e_t(\tau)$.
~\\

We are now in a position to obtain the PTA signals due to massive BH binaries inspiraling along 3PN-accurate eccentric orbits, and this is what we explore in the next subsection.

\subsection{Pictorial exploration of $R(t)$ due to BH binaries in relativistic eccentric orbits}
\label{sec:illustrated_results}

We begin by displaying temporally evolving quadrupolar order $h_{+,\times}^{Q}(t)$, specified by  Eqs.~(\ref{eq:h+x_N}), while employing our semi-analytic prescription for evolving $n$, $e_t$, $\gamma$, and $l$ in Fig.~\ref{fig:hpx_comparison}.
It should be noted that our explicit expressions for $h_{+,\times}^{Q}$ involve $u$ and therefore, we additionally need to invert the 3PN-accurate Kepler Equation, given by Eq.~(\ref{eq:KE_PN}), at every $l$ value to obtain the temporal evolution of our dominant order GW polarization states.
The treatment of PN-accurate Kepler Equation, as noted earlier, is performed by adapting 
and extending the Mikkola's method \cite{Mikkola87,TessmerGopakumar2007}. 
The resulting $h_{+,\times}(t)$ associated with massive BH binaries inspiraling along fully 3PN-accurate eccentric orbits are displayed in Fig.~\ref{fig:hpx_comparison}, and are labelled ``Conservative+Reactive''.
The effects of GW emission are clearly visible in $e=0.8$ plots and it causes certain 
{waveform dephasing}
while comparing with plots that do not include the effects of GW emission. 
Let us emphasize that our semi-analytic approach is capable of treating orbital eccentricities that are $\leq 1$ as we explicitly employ the eccentric anomaly $u$ to trace the PN-accurate eccentric orbit.

We now have all the ingredients to obtain ready-to-use PTA signals associated with  non-spinning SMBH binaries inspiraling along PN-accurate eccentric orbits. 
As mentioned earlier, the fact that $h_{+,\times}^{Q}$ expressions given by Eqs.~(\ref{eq:h+x_N}) explicitly contain $u$ and $\phi$ prevents us from evaluating analytically the integrals that appear in the expression for $R(t)$ as evident from Eqs.~(\ref{eq:R(t) int}-\ref{eq:spx(t) int}).
Therefore, we employ an adaptive numerical integration routine, namely the \texttt{QAG} routine \cite{Zwillinger1992}  to evaluate Eqs.~(\ref{eq:R(t) int}-\ref{eq:spx(t) int}) while computing pulsar timing residuals.
We first 
 provide a pictorial depiction of $R(t)$ and explain its various features with the help of $+/\times$ residual
 plots.

\begin{figure*}
\centering
\includegraphics[scale=0.57]{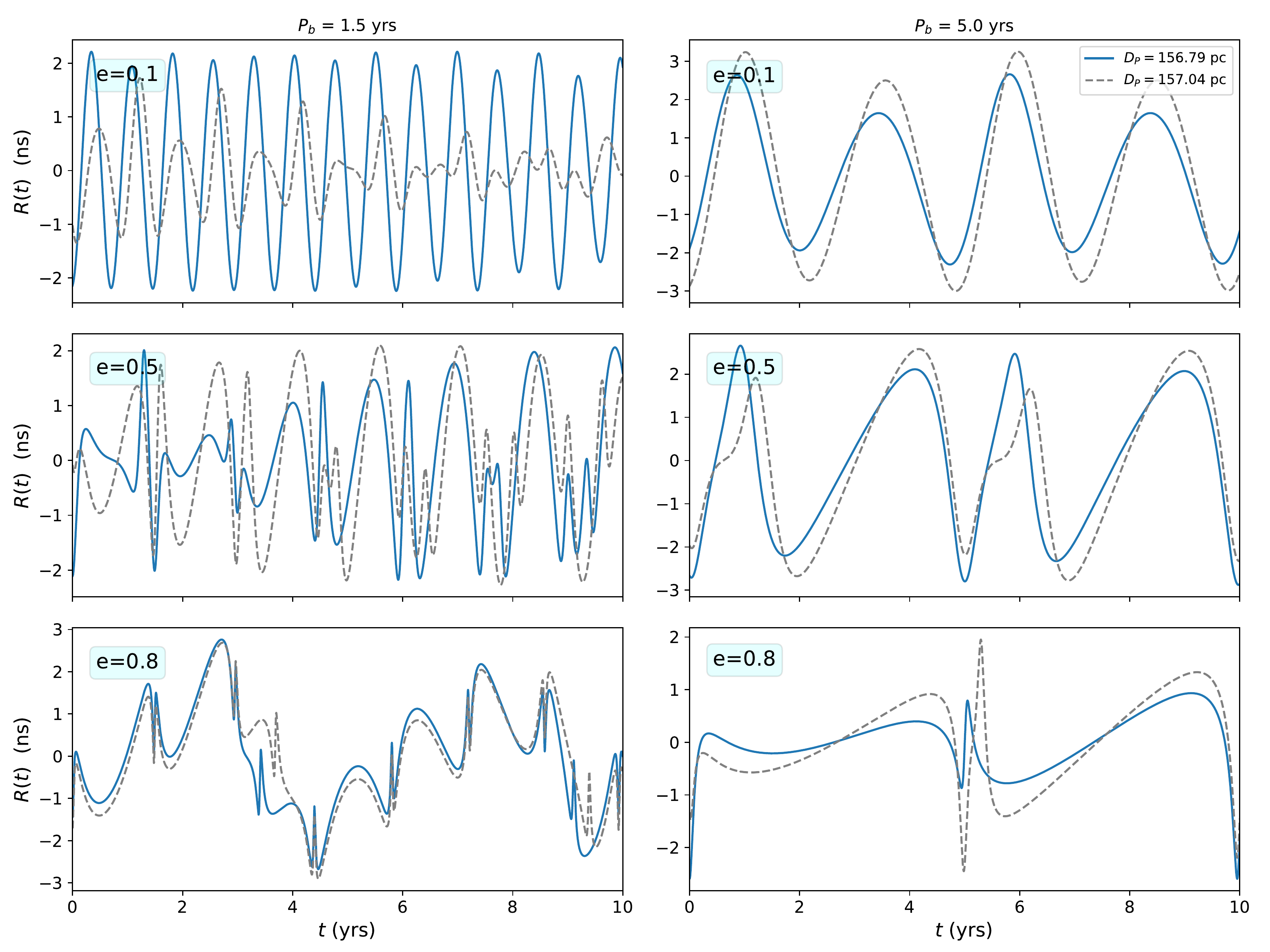}
\caption{
PTA signals induced on PSR J0437$-$4715 by a fiducial massive BH binary at $1$ Gpc away with two different $P_b$ values and three $e_t$ values at $t_E=0$ epoch.
The location of this pulsar is given by  RA $04^h37^m16^s$, DEC $-47^\circ 15' 09''$ and is at a distance of 156.79 pc \cite{Reardon2016}. 
The sky location of the binary is arbitrarily chosen to be RA $08^h00^m00^s$, DEC $-20^\circ 00'00''$ and other binary parameters are the same as in Fig.~\ref{fig:EccentricResiduals_Num_px}. 
The dashed plots correspond to a pulsar distance of 157.04 pc, which is off from the measured distance by its $1\sigma$ uncertainty \cite{Reardon2016}.
It is very clear that the features of $R(t)$ for $P_b=1.5$ yrs are very sensitive to the pulsar distance, especially for low and moderate eccentricities. This may be attributed to the frequencies of the Earth and pulsar terms being similar for low and moderate eccentricities  and the pulsar distance difference being roughly equivalent to half the orbital period.
In contrast, the features of the $P_b=5$ yrs case are much less sensitive to the pulsar distance as the pulsar distance difference is not tuned to the orbital period.
    }
    \label{fig:EccentricResiduals_Num_Both}
\end{figure*}

We display in Fig.~\ref{fig:EccentricResiduals_Num_Both}  PTA signals induced on  PSR J0437$-$4715 by a fiducial equal mass BH binary having $M=10^{9}\,M_{\astrosun}$ with face-on orbit ($i=0$) at a luminosity distance of $1$ Gpc, for three different eccentricities and two different orbital periods. 
We  let the sky location of the GW source to be RA $08^h00^m00^s$, DEC $-20^\circ00'00''$, with $\psi=0$. 
Each panel in Fig.~\ref{fig:EccentricResiduals_Num_Both} corresponds to a particular combination of orbital eccentricity and orbital period at the $t_E=0$ epoch.
Additionally, we choose two estimates for the pulsar distance, namely 156.79 pc and 157.04 pc, which are consistent with the $1\sigma$ uncertainty for its measurement, available in \rcite{Reardon2016}.
These choices lead to two plots each in six panels of  Fig.~\ref{fig:EccentricResiduals_Num_Both}.
Amplitude modulations, visible in the moderate to high eccentricity cases for $P_b=1.5$ yrs, are due to the fact that the pulsar term contributions can have substantially different orbital eccentricity and period for such high eccentric systems.
Interestingly, temporal evolution of $R(t)$ is pulsar distance-dependent especially for the lower and moderate $e$ values as evident from the first two panels for $P_b$=1.5 yrs.
Prominent dephasing in the $P_b$=1.5 yrs case may be due to the fact that the change in pulsar distance is roughly equivalent to half of the orbital period.
Such changes in the $R(t)$ evolution is less pronounced for the high $e$ case as the underlying frequencies of the  Earth and pulsar terms are significantly different.
In contrast, such strong dependence of $R(t)$ on the pulsar distance is not observed in the $P_b$=5 yrs case due to the fact that the pulsar distance difference is not tuned to the orbital period.
Interestingly, the epochs of the sharp features, visible in Fig.~\ref{fig:EccentricResiduals_Num_Both},  are very sensitive to the pulsar distance in the $P_b$=1.5 yrs case, and its implications are being investigated.

\begin{figure*}
    \centering
    \includegraphics[scale=0.59]{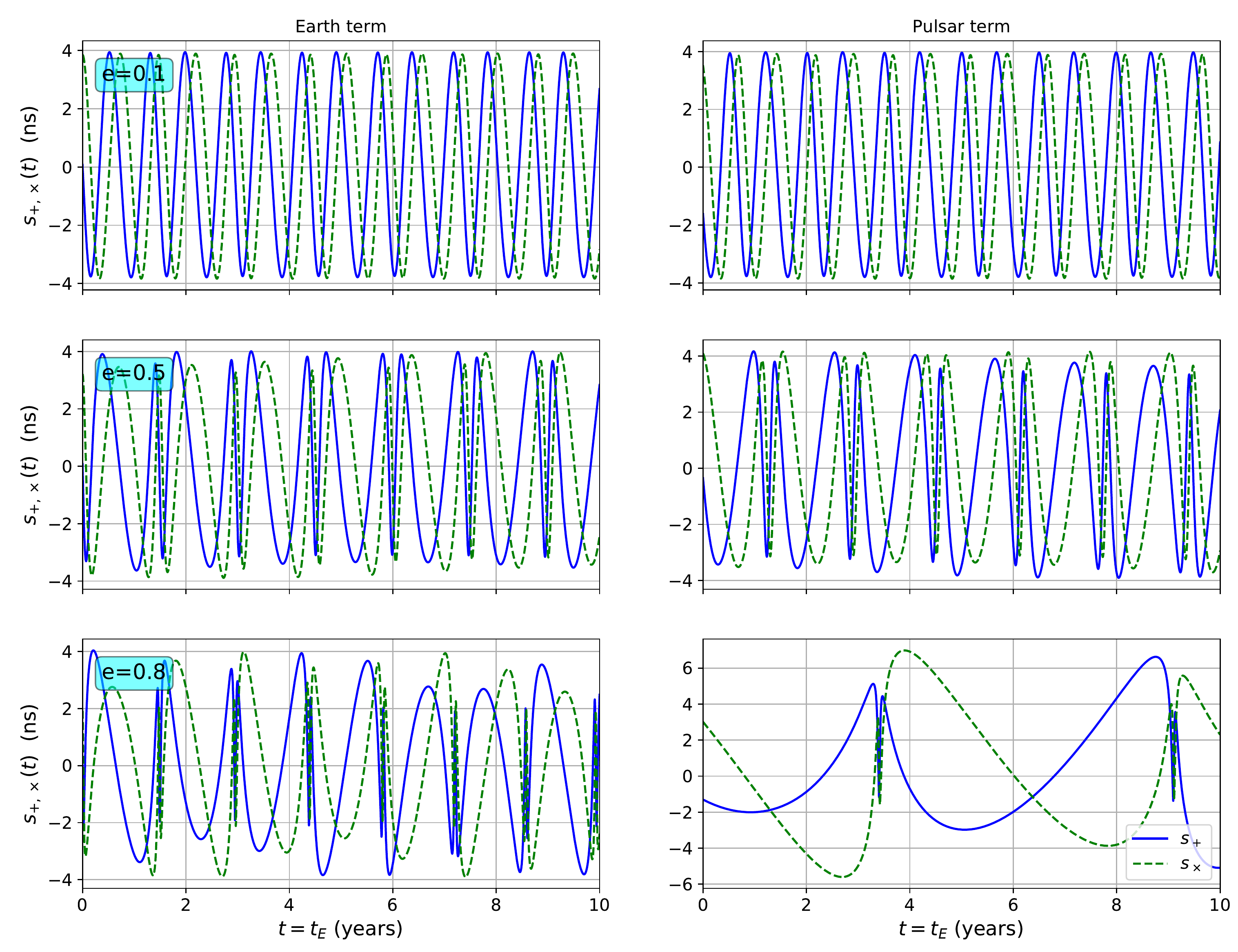}
    \caption{
Plots of $s_{+,\times}(t)$, namely 
the plus/cross  residuals, for a $1$ Gpc away equal mass binary BH having different orbital eccentricities at the 
initial Earth epoch while all other binary parameters are similar to those in Fig.~\ref{fig:hpx_comparison} with $P_b=1.5$ yrs at $t_E=0$.
We plot both the Earth and the Pulsar term contributions while assuming a geometric delay of $1000$ years between these two locations. 
The GW emission ensures that fiducial pulsar contributions to  $s_{+,\times}(t)$ have higher orbital eccentricities and periods. 
This is very prominent for the large     initial eccentricity ($e=0.8$) binary BH configuration.
}    
\label{fig:EccentricResiduals_Num_px}
\end{figure*}

To get a pulsar-independent view of these timing residuals, we plot in Fig.~\ref{fig:EccentricResiduals_Num_px} the associated $+/\times$ residuals while separating the Earth and the Pulsar term contributions using identical parameters to Fig.~\ref{fig:EccentricResiduals_Num_Both}, with $P_b=5$ yrs.
These plots confirm our earlier statement that the pulsar term, which provides a snapshot of the orbital configuration of our GW source  at an earlier epoch, can have substantially 
different orbital eccentricity and period, especially for highly eccentric BH binaries.
It is clearly the mixing of the two contributions with very different evolution timescales that produces various features present in our $R(t)$ plots.

\begin{figure}
    \centering
    \includegraphics[scale=0.57]{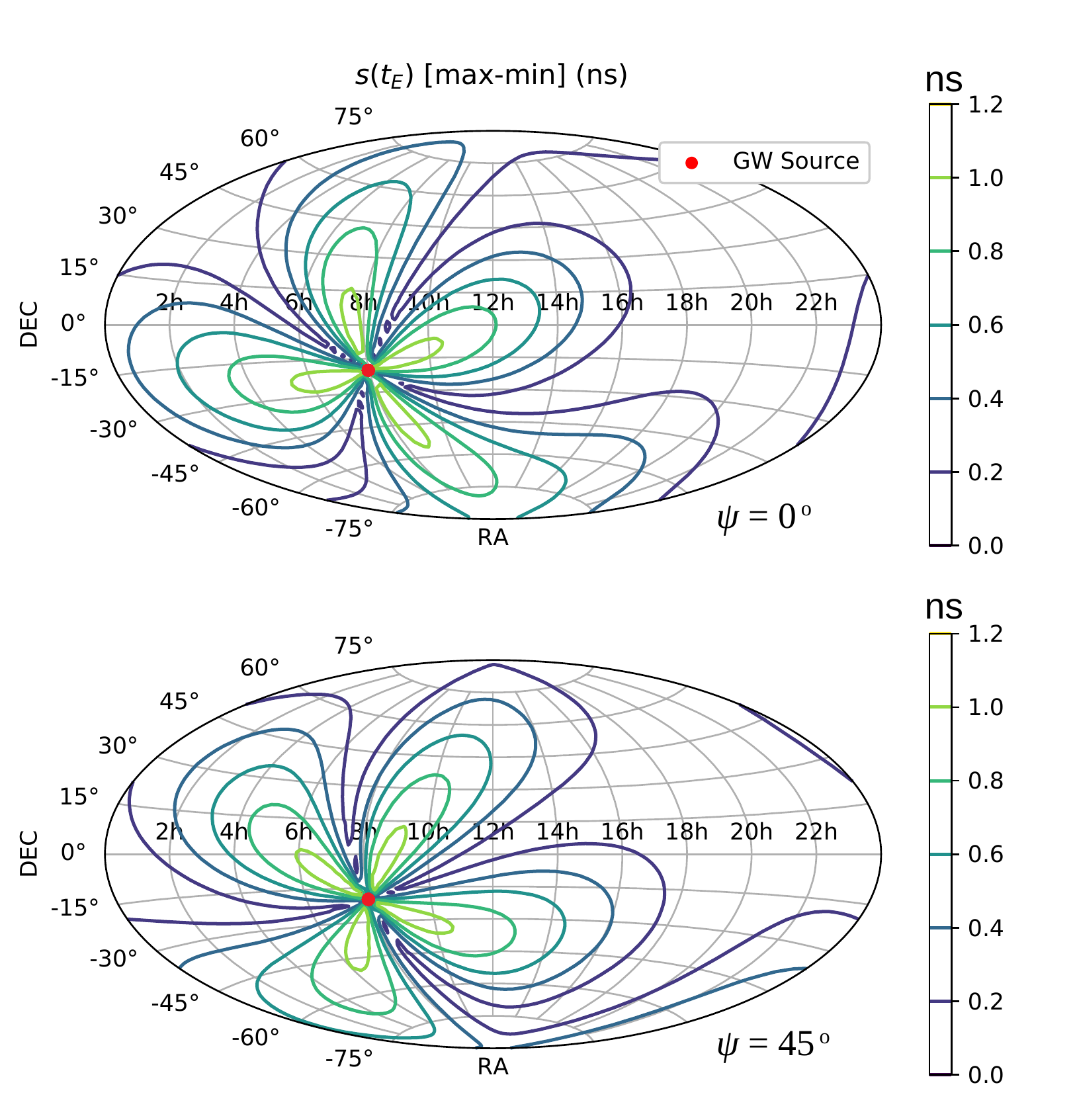}
    \caption{
    Graphical display of the `strength' of the Earth term $s(t_E)$ 
    as a function of pulsar's sky location. The two panels are for two specific values of 
    GW polarization angle and the rest of the parameters are identical to those 
    employed in Fig.~\ref{fig:EccentricResiduals_Num_Both} with $P_b=1.5$ yrs at $t_E=0$.
    In these plots, the red dot represents the sky location of the GW source and 
    the expected quadrupolar pattern is clearly visible. We rotate 
    the bottom panel plot by $45^\circ$ w.r.t the top panel plot in accordance with the $\psi$ values used.
    }
    \label{fig:sky_flower}
\end{figure}

We now proceed to display the quadrupolar nature of our PTA signal in 
Fig.~\ref{fig:sky_flower}.
Specifically, we plot certain strength of the Earth term 
as a function of the sky location of the pulsar for a given GW source.
This 
strength of the Earth term is defined  as the difference between the maximum and minimum of $s(t_E)$ within a given time span.
The top and bottom panels show  such $s(t_E)$ strength for $\psi=0$ and $\psi=45^\circ$ values, respectively. 
For these plots, we let the orbital 
 eccentricity of the GW source to be 0.5 and the all other parameters are same as in Fig.~\ref{fig:EccentricResiduals_Num_Both} and Fig.~\ref{fig:EccentricResiduals_Num_px}.
Our plots clearly show the quadrupolar pattern of the expected PTA signal, and the comparison between the top and bottom panels reveals the $45^\circ$ rotation that is expected from the $\psi$ values. 
Additionally, these plots essentially confirm that we are employing appropriate expressions 
for $F_\times$ and $F_+$.

~

We now turn our attention towards the numerical costs of our approach to obtain the temporal evolution of $n$, $e_t$, $\gamma$, and $l$ as well as the computation of the PTA signal $R(t)$, and this is what we explore in the next subsection.

\subsection{The cost of computing the  orbital evolution and the PTA signal}
\label{sec:computation_cost}
\begin{figure}
    \centering
    \includegraphics[scale=0.45]{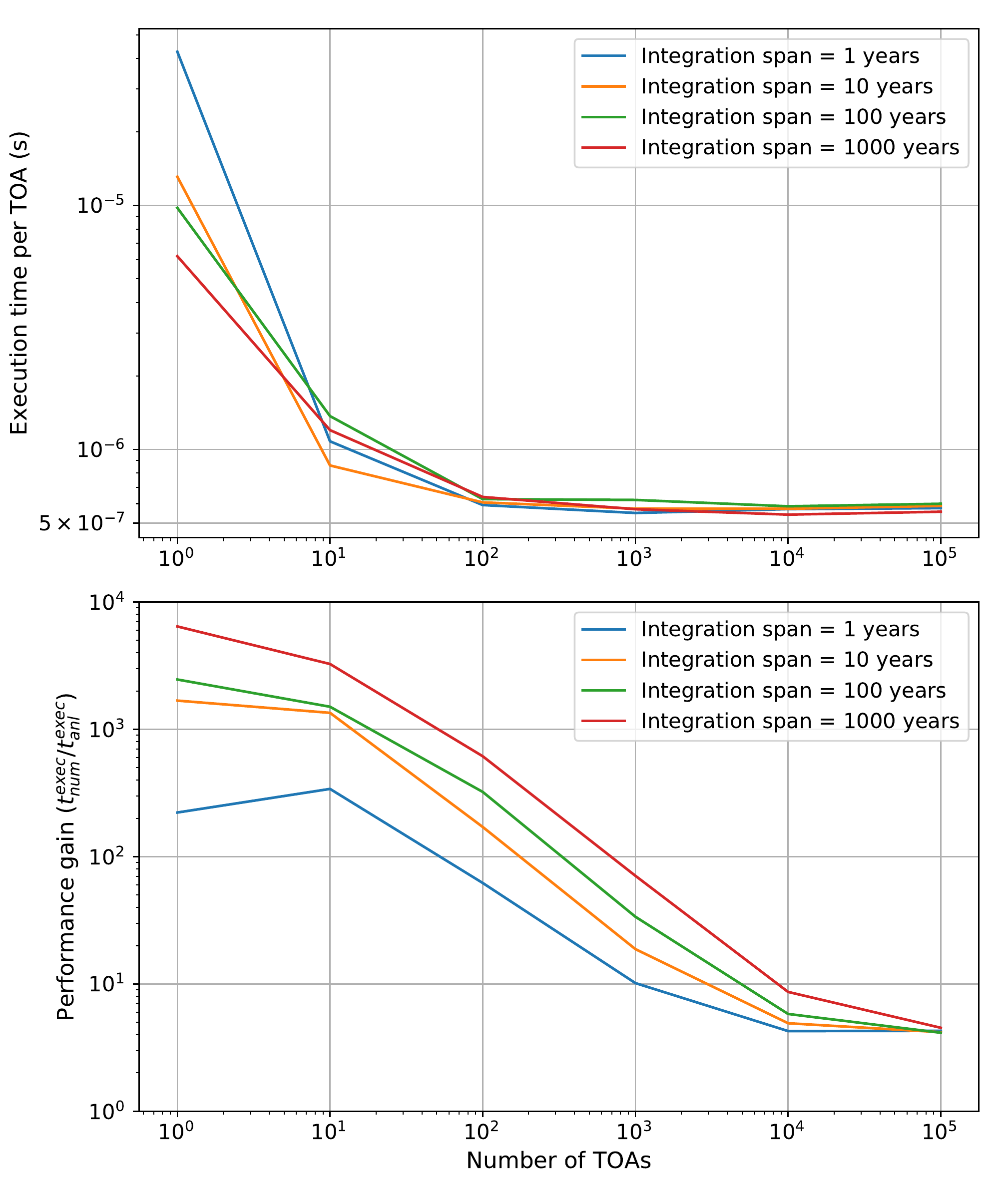}
    \caption{
 Plots that compare execution times associated  with the evolution of an eccentric binary  
using two approaches.
The numerical approach (\textit{num}) solves the set of four differential equations, given by Eqs.~(\ref{eq:ode_system}), by employing the \texttt{gsl\_odeiv2\_step\_rkf45} adaptive integration method of the GNU Scientific Library (\texttt{GSL}). 
The special functions, required by our analytic approach, are also evaluated with the help of \texttt{GSL}.
We consider equal mass BH binaries with $M=10^9\,M_{\astrosun}$, $P_{b0}=1.5$ years, $e_{t0}=0.5$, and let $l_0=\gamma_0=0$. 
These computations were performed in \texttt{C++} in an Intel Core i7 machine using a single core. These plots reveal that our semi-analytic approach is more efficient 
that the regular numerical approach.
}
    \label{fig:performance}
\end{figure}

We begin by comparing the computational cost of our semi-analytic approach against numerically solving Eqs.~(\ref{eq:ode_system}) to obtain the orbital evolution.
Excluding the one-time cost of computing the look-up table for $e(\tau)$, the execution time $t^\text{exec}$ taken to compute the state of the orbit $(n,e_t,\gamma,l)$ at a given set of TOAs should depend on the number of TOAs ($N_\text{TOA}$) as well as their total observation span/integration span (specified by some $t_0$ and $t_1$).
This is illustrated in the top panel of Fig.~\ref{fig:performance} where we plot the execution time per TOA ($t^\text{exec}/N_\text{TOA}$) required to compute our variables $(n,e,\gamma,l)$ as a function of $N_\text{TOA}$ for different integration spans ($t_1-t_0$) in our semi-analytic approach. 
This panel shows that the computational time required for evaluating $(n,e,\gamma,l)$ at a given TOA is independent of both the integration span as well as the number of TOAs when the number of TOAs is sufficiently large.  
This impressive feature may be contrasted  with the fact that the execution time, when TOA numbers are small, is dominated by the one-time evaluation of various coefficients like $\kappa$, $\alpha$ and $\beta$.

The bottom panel of Fig.~\ref{fig:performance} compares the performance gain of our semi-analytic method with respect to the usual approach of solving numerically Eqs.~(\ref{eq:ode_system})
by employing the ratio of execution times ($t_{\text{num}}^{\text{exec}}/t_{\text{anl}}^{\text{exec}}$).
The associated plots reveal that this ratio 
 increases substantially as one increases the  integration span, especially for low $N_{\text{TOA}}$ values.
 However, the ratio eventually 
  decreases and essentially converges to a value close to $5$ when  $N_{\text{TOA}}$ is a large number.
This behavior is expected, since a numerical {\it ODE solver} is  required to compute the right hand side of Eqs.~(\ref{eq:ode_system}) at many points between the TOAs where the solutions are required while evolving the binary over time.
In contrast, our semi-analytic approach {\it only  computes} the solutions  at the required TOAs. 
However, as the number of TOAs within an integration span increases, the number of intermediate points required by the numerical solver decreases too.
This leads to the  behavior displayed in the bottom panel of Fig.~\ref{fig:performance}, and we infer that the semi-analytic solution usually outperforms the numerical one.

\begin{figure}
    \centering
    \includegraphics[scale=0.45]{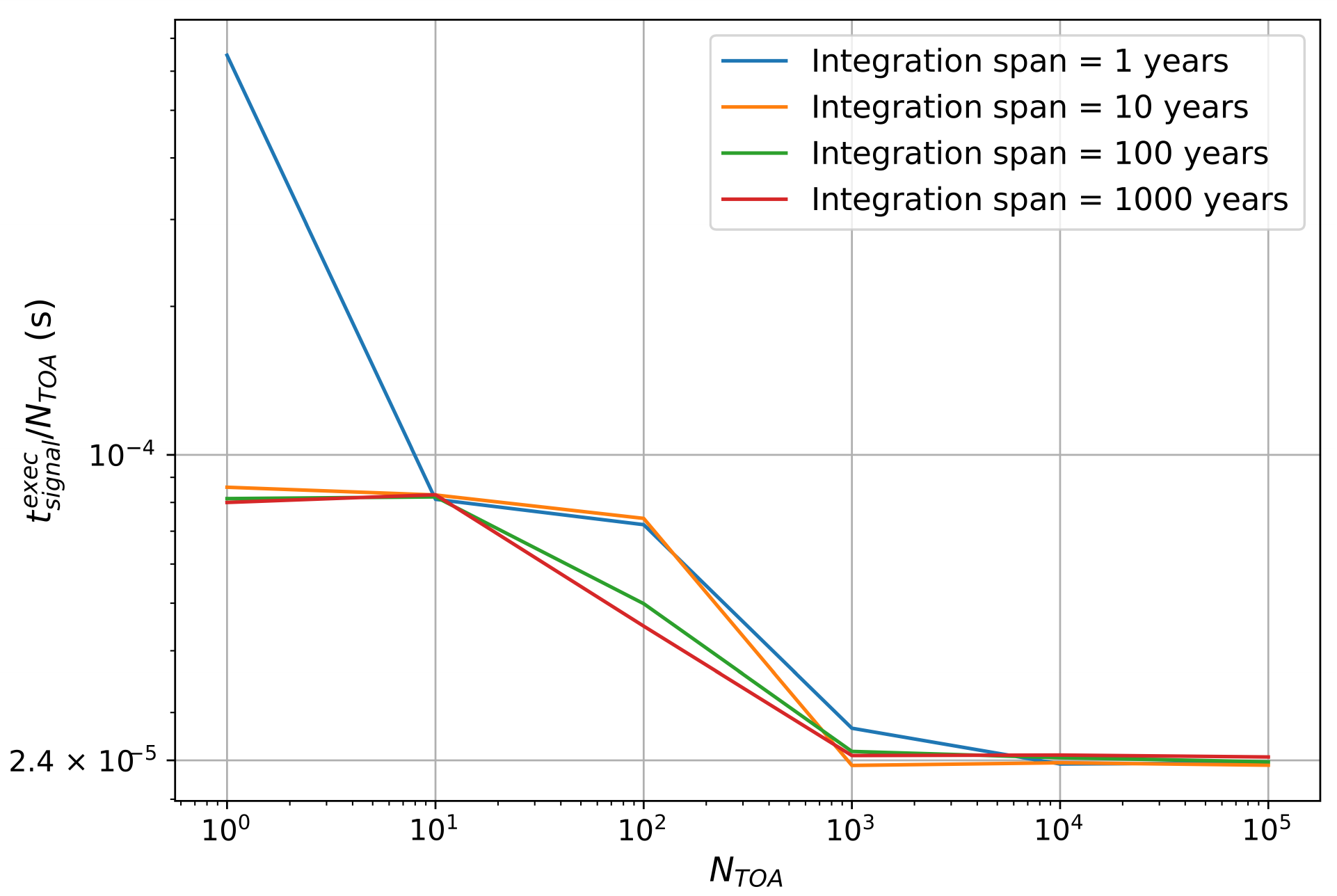}
    \caption{The execution time per TOA for computing $R(t)$. For small $N_\text{TOA}$ the execution time is dominated by one-time computations whereas for large $N_\text{TOA}$ it is essentially independent of $N_\text{TOA}$. A comparison with the top panel of Fig.~\ref{fig:performance} also reveals that the execution time is dominated by the numerical integration of $h(t)$ to compute $R(t)$.
    }
    \label{fig:R(t)_perf}
\end{figure}

Fig.~\ref{fig:R(t)_perf} shows the time taken to compute the PTA signal $R(t)$ per TOA ($t^\text{exec}_\text{signal}/N_\text{TOA}$) as a function of $N_\text{TOA}$ for different integration spans. 
Once again, we see that the execution time is dominated by the one-time computations when $N_\text{TOA}$ is small, but is independent of $N_\text{TOA}$ when $N_\text{TOA}$ is large. 
A comparison of Fig.~\ref{fig:R(t)_perf} with the top panel of Fig.~\ref{fig:performance} reveals that  the execution time of computing $R(t)$ is dominated by the cost of numerically integrating $h(t)$ to get $R(t)$.

~

Clearly, it is desirable to provide appropriate checks to verify the correctness of
our detailed prescription for computing  pulsar timing residuals, induced by relativistic eccentric binaries
as it involves many numerical ingredients and detailed and lengthy analytic expressions.
This is pursued in the next section where we provide a fully analytic way to compute $+/\times$ residuals
for non-spinning BH binaries moving in PN-accurate \textit{moderately eccentric} orbits.

\section{Fully analytic $+/\times$ residuals for binaries in Post-Keplerian small-eccentricity orbits}
\label{sec:analytic_residuals}

This section provides a fully analytic way of computing pulsar timing residuals due to BH binaries moving in quasi-Keplerian orbits of moderate eccentricities. 
This effort invokes explicit analytic expressions for $h_{+,\times}^{Q}(t)$ that are associated with non-spinning compact binaries moving in conservative 3PN-accurate small eccentricity orbits, derived in \rcite{Boetzel2017}. 
The main motivation, as noted earlier, is to provide a powerful check on the results, originating from our semi-analytic approach, for computing  $s_{+,\times}(t)$ 
associated with 
quasi-Keplerian orbits of arbitrary  eccentricities. 
The present section is also influenced by \rcite{Taylor2016} that provided explicit analytic expressions for the quadrupolar order $+,\times$ residuals from BH binaries in Newtonian eccentric orbits.
 
 The effort, detailed in \rcite{Taylor2016} , employs various results 
from the Fourier analysis of the classical Kepler equation in terms of the Bessel functions, available in \rcite{Colwell1993} and  apply them in the quadrupolar order $h_{+,\times}^{Q}$ expressions, given by Eqs.~(\ref{eq:h+x_N}) \cite{PetersMathews1693}.
The resulting fully analytic Newtonian GW polarization states may be symbolically written as \cite{Taylor2016}
\begin{subequations}
\begin{align}
    h_{+}(t)&=\sum_{p=0}^{\infty}\left(a_{p}^{+}\cos\left(pl\right)\cos\left(2\omega\right)\right.\nonumber\\&\quad\quad\quad\left.+b_{p}^{+}\sin\left(pl\right)\sin\left(2\omega\right)+c_{p}^{+}\cos\left(pl\right)\right)\,,\\
    h_{\times}(t)&=\sum_{p=0}^{\infty}\left(a_{p}^{\times}\cos\left(pl\right)\sin\left(2\omega\right)\right.\nonumber\\&\quad\quad\quad\left.+b_{p}^{\times}\sin\left(pl\right)\cos\left(2\omega\right)\right)\,,
\end{align}
\label{eq:PM_waveform}
\end{subequations}
where the coefficients $a_p^{+,\times}$, $b_p^{+,\times}$ and $c_p^{+}$ contain trigonometric functions of the orbital inclination  $i$, 
while the orbital eccentricity $e_t$ enters in terms of Bessel functions of the first kind \cite{Moreno-Garrido1995}.
Recall that $\omega$ provides the argument of periapsis, which remains a constant for Newtonian orbits.
This ensures that such Newtonian compact binaries emit GWs at frequencies that are integer harmonics of $n$.
It is also possible to incorporate in an ad-hoc manner the linear-in-time evolution of $\omega$  to the above Newtonian order $h_{+,\times}$ expressions \cite{Seto2001,BarackCutler2004}.  
Employing the above Newtonian order expressions for the two  GW polarizations states,  \rcite{Taylor2016} computed analytically the $+/\times$ residuals which may be written symbolically as 
\begin{subequations}
\begin{align}
  s_{+}(t)&=\sum_{p=0}^{\infty}\frac{1}{np}\left(a_{p}^{+}\sin\left(pl\right)\cos\left(2\omega\right)\right.\nonumber\\&\quad\quad\quad\left.-b_{p}^{+}\cos\left(pl\right)\sin\left(2\omega\right)+c_{p}^{+}\sin\left(pl\right)\right)\,,\\
  s_{\times}(t)&=\sum_{p=0}^{\infty}\frac{1}{np}\left(a_{p}^{\times}\sin\left(pl\right)\sin\left(2\omega\right)\right.\nonumber\\&\quad\quad\quad\quad\left.-b_{p}^{\times}\cos\left(pl\right)\cos\left(2\omega\right)\right)\,.
\end{align}
\label{eq:PM_Residuals}
\end{subequations}
The explicit form of these coefficients may be easily extracted with the help of Eqs.~(21) and (22) of \rcite{Taylor2016}.
In what follows, we provide a fully post-Newtonian accurate version of these results.

Recall that 
fully analytic $h_{+,\times}^{Q}(l)$ expressions for compact binaries moving in conservative 3PN-accurate quasi-Keplerian small eccentric orbits were derived in \rcite{Boetzel2017}. 
This derivation  employed Eqs.~(\ref{eq:h+x_N}) for $h_{+,\times}^{Q}$ and an analytic treatment of the PN-accurate Kepler equation.
The detailed analysis of \rcite{Boetzel2017} provided PN-accurate expressions for both eccentric and true anomalies in terms of  infinite series expressions involving 
$l$ and $e_t$.
We write symbolically  the resulting quadrupolar order $h_{+,\times}(l)$ expressions as 
\begin{align}
    h_{+,\times}(t)&=\sum_{p=0}^{\infty}\sum_{q=0}^{\infty}\left\{ a_{p,q}^{+,\times}\cos\left(pl\right)\cos\left(q\lambda\right)\right.\nonumber\\&\quad\quad\left.+b_{p,q}^{+,\times}\sin\left(pl\right)\cos\left(q\lambda\right)\nonumber\right.\\&\quad\quad\left.+c_{p,q}^{+,\times}\cos\left(pl\right)\sin\left(q\lambda\right)\nonumber\right.\\&\quad\quad\left.+d_{p,q}^{+,\times}\sin\left(pl\right)\sin\left(q\lambda\right)\,\right\} \,,
    \label{eq:BSGKJ_waveform}
\end{align}
where we have defined $\lambda=l+\gamma$ \cite{Boetzel2017}.
A straightforward integration of the above expression leads to
\begin{align}
    s_{+,\times}(t)&=\frac{1}{n}\sum_{p,q=0}^{\infty}{\vphantom{\sum}}'\left\{ A_{p,q}^{+,\times}\cos\left(pl\right)\cos\left(q\lambda\right)\right.\nonumber\\&\qquad\left.+B_{p,q}^{+,\times}\sin\left(pl\right)\cos\left(q\lambda\right)\right.\nonumber\\&\qquad\left.+C_{p,q}^{+,\times}\cos\left(pl\right)\sin\left(q\lambda\right)\right.\nonumber\\&\qquad\left.+D_{p,q}^{+,\times}\sin\left(pl\right)\sin\left(q\lambda\right)\,\right\} \,,
    \label{eq:BSGKJ_residuals}
\end{align}
where we have ignored the effects of GW emission while performing various integrations. 
This is justified as the radiation reaction timescale is substantially larger than the orbital and advance of periapsis timescales. 
Further, the primed sum excludes the $p=q=0$ term in the above expressions. 
These multi-index  $A$, $B$, $C$, $D$ coefficients
involve $x$, $\eta$, trigonometric functions of $i$, and $e_t$ contributions via infinite series of Bessel functions. 
They may be expressed as 
\begin{subequations}
\begin{align}
  A_{p,q}^{+,\times}&=\frac{-p\,b_{p,q}^{+,\times}+(1+k)q\,c_{p,q}^{+,\times}}{p^{2}-(1+k)^{2}q^{2}}\,,\\
  B_{p,q}^{+,\times}&=\frac{p\,a_{p,q}^{+,\times}+(1+k)q\,d_{p,q}^{+,\times}}{p^{2}-(1+k)^{2}q^{2}}\,,\\
  C_{p,q}^{+,\times}&=\frac{-p\,d_{p,q}^{+,\times}-(1+k)q\,a_{p,q}^{+,\times}}{p^{2}-(1+k)^{2}q^{2}}\,,\\
  D_{p,q}^{+,\times}&=\frac{p\,c_{p,q}^{+,\times}-(1+k)q\,b_{p,q}^{+,\times}}{p^{2}-(1+k)^{2}q^{2}}\,.
\end{align}
\label{eq:BSGKJ_residual_coeffs}
\end{subequations}

\begin{figure*}
    \centering
    \includegraphics[scale=0.55]{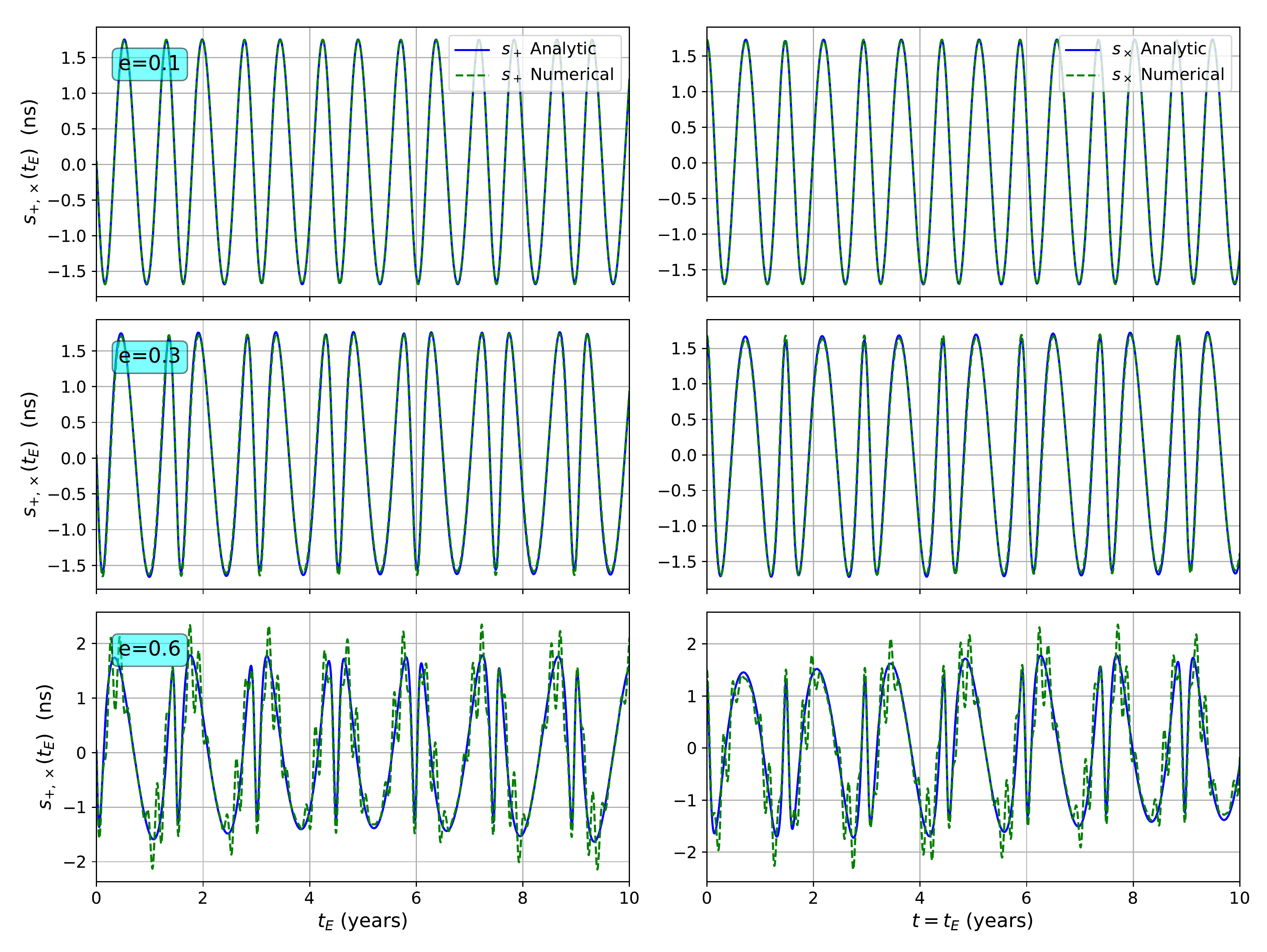}
    \caption{
    We overlay $s_{+,\times}(t)$  plots 
    for an equal mass binary with different eccentricities  that arise from our fully
    analytic post-circular approximation and the earlier described semi-analytic apporach.
    The binary parameters are the same as in Figure \ref{fig:EccentricResiduals_Num_px}  with $P_b=1.5$ yrs at $t_E=0$ and 
    the post-circular approximation includes all terms accurate up to $\mathcal{O}(e_t^8)$. 
    We observe that the quality of the post-circular approximation degrades
      as the eccentricity increases, as expected. 
    }
    \label{fig:EccentricResiduals_Anl_eg}
\end{figure*}

Clearly, it is neither advisable nor feasible to evaluate these coefficients for arbitrarily high $p$ and $q$ values  to high precision. 
This is because the underlying Bessel function evaluations are computationally  very expensive. 
However, it is straightforward to obtain Taylor expansions of these coefficients around $e_t=0$.
The resulting expansions, accurate up to some $\mathcal{O}(e_t^m)$, ensure that the Fourier coefficients beyond a certain $p_{\max}$ and $q_{\max}$ vanish for any given $m$. 
This is essentially due to the following property of the Bessel functions of the first kind
\[
    \lim_{x\rightarrow 0}\text{J}_m(x) \sim \mathcal{O}(x^m)\,. 
\]
Unfortunately, both the Fourier series, given by Eqs.~(\ref{eq:BSGKJ_residual_coeffs}) and the associated power series expansions for the involved $A$, $B$, $C$, $D$ coefficients converge slowly for moderately large $e_t$ values.
This might signal the breaking down of the approximation and may be associated with the celebrated Laplace limit \cite{Watson1966}.
Detailed comparisons of various Bessel function contributions, computed numerically and analytically, reveal that 
such an expansion accurate up to $\mathcal{O}(e_t^8)$ can be used to compute timing residuals for eccentricities less than $0.3$. 
In what follows, we display the explicit expressions for the quadrupolar order $h^{Q}_{+}(l)$ that includes all the
eccentricity corrections up to $\mathcal{O}(e_t^4)$, and the associated $+$ residual:
\begin{widetext}
{\small
\begin{subequations}
\begin{align}
h_{+}^{Q}&={\cal H}_{0}\Bigg[\left(e_{t}s_{i}^{2}-\frac{1}{8}e_{t}^{3}s_{i}^{2}\right)\cos(l)+\left(e_{t}^{2}s_{i}^{2}-\frac{1}{3}e_{t}^{4}s_{i}^{2}\right)\cos(2l)\nonumber\\
&\quad+\left(-\frac{1}{8}23e_{t}^{4}c_{i}^{2}+5e_{t}^{2}c_{i}^{2}-2c_{i}^{2}-\frac{23e_{t}^{4}}{8}+5e_{t}^{2}-2\right)\cos(2\lambda)\nonumber\\
&\quad+\left(\frac{1}{8}e_{t}^{4}c_{i}^{2}+\frac{e_{t}^{4}}{8}\right)\cos(4l-2\lambda)+\left(-\frac{1}{4}81e_{t}^{4}c_{i}^{2}-\frac{81e_{t}^{4}}{4}\right)\cos(2\lambda+4l)\nonumber\\
&\quad+\left(\frac{171}{16}e_{t}^{3}c_{i}^{2}-\frac{9e_{t}c_{i}^{2}}{2}+\frac{171e_{t}^{3}}{16}-\frac{9e_{t}}{2}\right)\cos(2\lambda+l)+\left(\frac{7}{48}e_{t}^{3}c_{i}^{2}+\frac{7e_{t}^{3}}{48}\right)\cos(3l-2\lambda)\nonumber\\
&\quad+\left(-\frac{1}{48}625e_{t}^{3}c_{i}^{2}-\frac{625e_{t}^{3}}{48}\right)\cos(2\lambda+3l)
+\left(20e_{t}^{4}c_{i}^{2}-8e_{t}^{2}c_{i}^{2}+20e_{t}^{4}-8e_{t}^{2}\right)\cos(2\lambda+2l)\nonumber\\
&\quad+\left(-\frac{1}{16}13e_{t}^{3}c_{i}^{2}+\frac{3e_{t}c_{i}^{2}}{2}-\frac{13e_{t}^{3}}{16}+\frac{3e_{t}}{2}\right)\cos(l-2\lambda)
+\frac{9}{8}e_{t}^{3}s_{i}^{2}\cos(3l)+\frac{4}{3}e_{t}^{4}s_{i}^{2}\cos(4l)\Bigg]\,,
\end{align}
\begin{align}
s_{+}^{Q}&=\frac{{\cal H}_{0}}{n}\Bigg[\;\left(e_{t}s_{i}^{2}-\frac{1}{8}e_{t}^{3}s_{i}^{2}\right)\sin(l)+\left(\frac{1}{2}e_{t}^{2}s_{i}^{2}-\frac{1}{6}e_{t}^{4}s_{i}^{2}\right)\sin(2l)\nonumber\\
&\quad+\left(-\frac{1}{16}23e_{t}^{4}c_{i}^{2}+\frac{5}{2}e_{t}^{2}c_{i}^{2}-c_{i}^{2}-\frac{23e_{t}^{4}}{16}+\frac{5e_{t}^{2}}{2}-1\right)\sin(2\lambda)\nonumber\\
&\quad+\left(\frac{13}{16}e_{t}^{3}c_{i}^{2}-\frac{3e_{t}c_{i}^{2}}{2}+\frac{13e_{t}^{3}}{16}-\frac{3e_{t}}{2}\right)\sin(l-2\lambda)
+\left(\frac{57}{16}e_{t}^{3}c_{i}^{2}-\frac{3e_{t}c_{i}^{2}}{2}+\frac{57e_{t}^{3}}{16}-\frac{3e_{t}}{2}\right)\sin(2\lambda+l)\nonumber\\
&\quad+\left(\frac{1}{16}e_{t}^{4}c_{i}^{2}+\frac{e_{t}^{4}}{16}\right)\sin(4l-2\lambda)+\left(-\frac{1}{8}27e_{t}^{4}c_{i}^{2}-\frac{27e_{t}^{4}}{8}\right)\sin(2\lambda+4l)\nonumber\\
&\quad+\left(\frac{7}{48}e_{t}^{3}c_{i}^{2}+\frac{7e_{t}^{3}}{48}\right)\sin(3l-2\lambda)+\left(-\frac{1}{48}125e_{t}^{3}c_{i}^{2}-\frac{125e_{t}^{3}}{48}\right)\sin(2\lambda+3l)\nonumber\\
&\quad+\left(5e_{t}^{4}c_{i}^{2}-2e_{t}^{2}c_{i}^{2}+5e_{t}^{4}-2e_{t}^{2}\right)\sin(2\lambda+2l)
+\frac{1}{3}e_{t}^{4}s_{i}^{2}\sin(4l)+\frac{3}{8}e_{t}^{3}s_{i}^{2}\sin(3l)\;\Bigg]\,,
\end{align}
\end{subequations}
}
\end{widetext}
where we have defined ${\mathcal{H}}_0 = \frac{GM\eta}{D_L c^2}x$.
We note in passing that we have explicitly computed the quadrupolar order $h_{+,\times}(l)$ and its temporally evolving $+/\times$ residuals that  include
all the $\mathcal{O}(e_t^8)$ corrections. Additionally, these expressions were employed while making comparisons of our analytic and semi-analytic approaches to compute $s_{+,\times}(t)$, displayed in Fig.~\ref{fig:EccentricResiduals_Anl_eg}.

\begin{figure*}
    \centering
    \includegraphics[scale=0.5]{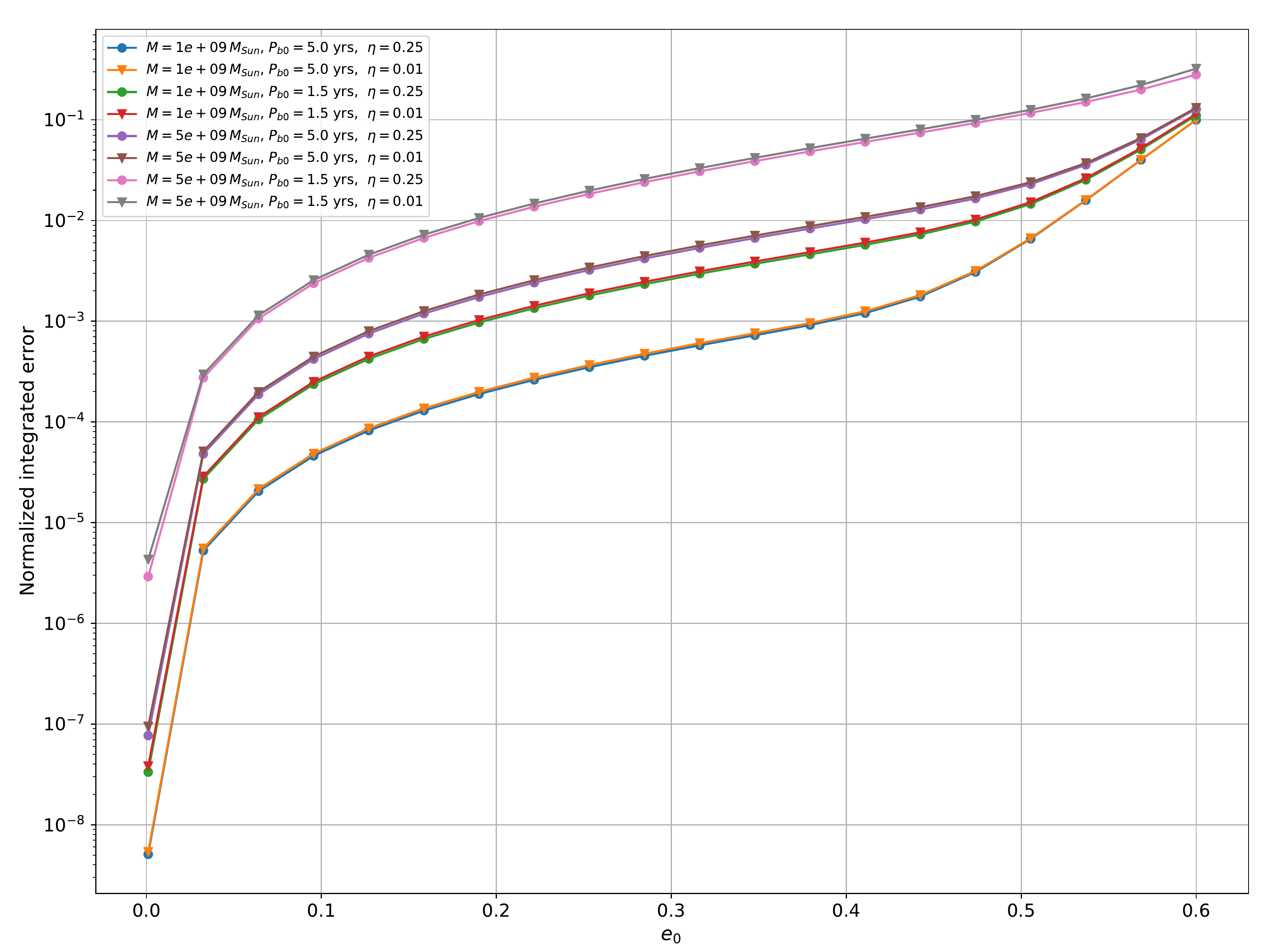}
    \caption{The normalized integrated error in the post-circular approximation as a function of the initial eccentricity for different combinations of $P_b$, $M$ and $\eta$. Rest of the binary parameters are identical to those used in Fig.~\ref{fig:EccentricResiduals_Anl_eg} and the normalized integrated error is defined using  Eq.~(\ref{eq:norm_err}). Clearly, the approximation gets progressively worse as we increase initial eccentricity values. In addition, the accuracy of the approximation decreases with shorter orbital periods and with higher masses. However, error is only weakly dependent on the mass ratio.}
    \label{fig:anl_mismatch}
\end{figure*}

We are now in a position to use these expressions to test our involved semi-analytical approach to obtain $+/\times$ residuals valid for arbitrary eccentricities. 
In Fig.~\ref{fig:EccentricResiduals_Anl_eg}, we overlay plots of $s_{+,\times}^{Q}(t)$ that arise from the above mentioned analytic approach and our semi-analytic approach while focusing only on the Earth term for three initial values of $e_t$. 
Additionally, we let the orbital elements and angles to vary according to our improvised GW phasing approach, detailed in Sec.~\ref{sec:orbital_evolution},  in both the approaches.
We observe excellent agreement between the two approaches for initial $e_t$ values up to
$0.3$ and  it is difficult to distinguish the dashed line plots in the first two panels. Therefore, 
these plots give us the confidence about the correctness of our semi-analytic approach to obtain $R(t)$ for BH binaries  inspiraling along relativistic eccentric orbits.
However, our analytic post-circular approach becomes progressively worse for a larger 
initial $e$ value as evident from the bottom panel plots.
We quantify the deviation between our semi-analytic and fully analytic temporally 
evolving plus/cross residuals with the help of the following 
 normalized integrated error defined as
{\small
\begin{equation}
    \varepsilon(e_0) = \frac{\sum_{i}\left[\left(s_{+}^{\text{num}}(t_{i})-s_{+}^{\text{anl}}(t_{i})\right)^{2}+\left(s_{\times}^{\text{num}}(t_{i})-s_{\times}^{\text{anl}}(t_{i})\right)^{2}\right]}{\sum_{i}\left[\left(s_{+}^{\text{num}}(t_{i})\right)^{2}+\left(s_{\times}^{\text{num}}(t_{i})\right)^{2}\right]}\,.
    \label{eq:norm_err}
\end{equation}}
In Fig.~\ref{fig:anl_mismatch}, we plot $\varepsilon$ as a function of initial orbital eccentricity for different combinations of $P_b$, $M$ and $\eta$.
This plot reinforces our conclusion that our post-circular approximation shows good agreement with the numerical approach for  $e<0.3$ values. 
The accuracy of the post-circular approximation is also seen to degrade for shorter orbital periods and for higher masses (i.e., more relativistic).
This behaviour is reflective of the truncation error arising from the analytic Fourier series solution for the 3PN Kepler equation and it is discussed in detail in \rcite{Boetzel2017}.
We note in passing that substantial differences between our two approaches for higher $e_0$ values may be related to the Laplace limit associated with the analytic solution to the classical Kepler equation.

\section{Summary and Discussions }
\label{sec:summary}
The present work provides a computationally efficient way to compute pulsar timing residuals induced by GWs from isolated massive BH binaries inspiraling along general relativistic eccentric orbits.
The use of an improvised version of the GW phasing approach, detailed in \rcite{Damour2006,KonigsdorfferGopakumar2006}, and the PN-accurate quasi-Keplerian parametrization
allowed us to model binary BH orbits that inspiral due to the emission of quadrupolar GWs along 3PN-accurate eccentric orbits in an essentially analytic manner.
This leads to analytic solutions for the mean motion $n$, mean anomaly $l$ and the periapsis angle $\gamma$ in terms of PN-accurate time eccentricity $e_t$ as well as system-dependent constants and initial conditions. 
This is augmented by using a computationally efficient way to obtain certain scaled temporal evolution for $e_t$ imposed by the quadrupolar order GW emission.
These inputs allowed us to obtain the quadrupolar order temporally evolving GW polarization states, the associated $+/{\times}$ residuals and the resulting pulsar timing residuals $R(t)$ due to PN-accurate eccentric inspirals in a computationally inexpensive way.
Additionally, we provided a fully analytic prescription to compute anaytic $+/{\times}$ residuals due to BH binaries moving in 3PN-accurate small eccentricity orbits.  
The excellent agreement between these two approaches provided a powerful check for our  very involved semi-analytic approach, appropriate for arbitrary orbital eccentricities.

We have implemented our prescription to compute pulsar timing residuals induced by GWs from arbitrary eccentricity BH binaries, developed in Sec. \ref{sec:EccentricResiduals}, as well as our fully analytic prescription to compute timing residuals for low-eccentricity binaries developed in Sec. \ref{sec:analytic_residuals}, in a \texttt{C++} package called \texttt{GWecc}
\footnote{\texttt{GWecc}, together with a Python wrapper, are available at \url{https://github.com/abhisrkckl/GWecc}.}. 
We are working to integrate these codes into the popular PTA-relevant packages like \texttt{TEMPO2} and \texttt{Enterprise}. 
This should allow us to constrain the presence of isolated eccentric BH binaries in 
the latest Parkes Pulsar Timing Array (PPTA) dataset \cite{SusobhananInPrep_PPTA}.
Further, efforts are on-going to tackle the IPTA DR2 and Nanograv 12.5 year  datasets by employing the present prescription \cite{CheeseboroInPrep_nanograv}.  
Clearly, it will be interesting to explore the effects of higher order GW radiation reaction effects in the equations for $ \dot n$ and $\dot e_t$. It is reasonable to expect that such contributions will be more relevant for the pulsar contributions to $R(t)$ due to the lengthy temporal separation between the Earth and the Pulsar epochs and this is currently under investigation.
Moreover, we are also pursuing detailed investigations 
on the implementation of certain 
Generalized Likelihood Ratio Test for the PTA detection of eccentric precessing BH binaries, influenced by \rcite{Wang2015}.

It turns out that the spin-orbit coupling can influence the nature of PTA signals from non-spinning massive BH binaries as this contribution  enters the dynamics at the 1.5PN order.
Therefore, we are extending the present approach by incorporating the spin effects, influenced by \rcite{Mingarelli2012,ChenYang2018}.
This effort relies on the availability of a Keplerian-type parametric solution for the dynamics of compact binaries that incorporates the effects of dominant order spin-orbit interactions \cite{KonigsdorfferGopakumar2005}. 

\begin{acknowledgments}
We thank Yannick Boetzel for helpful discussions and providing his \texttt{Mathematica} notebook, Lankeswar Dey and Xingjiang Zhu for helpful discussions, and Belinda Cheeseboro for testing and providing valuable suggestions regarding \texttt{GWecc}. AS wishes to thank the hospitality of CSIRO Astronomy and Space Sciences, ARC Centre of Excellence for Gravitational Wave Discovery (OzGrav) and Monash University. AS was partially supported by CSIRO and the Sarojini Damodaran Fellowship during the course of the collaboration.
AS \& AG acknowledge support of the Department of Atomic Energy, Government of India, under project \# 12-R\&D-TFR-5.02-0200.
S.R.T. acknowledges support from the NANOGrav project, which is supported by the National Science Foundation (NSF) Physics Frontier Center award number 1430284.
\end{acknowledgments}

\appendix

\begin{widetext}

\section{Higher order PN corrections to $\gamma(t)$}
\label{sec:gamma_PN}

This appendix details our approach to integrate 3PN-accurate expression for $d \gamma/dt$ which 
may be written symbolically as 
\begin{equation}
    \frac{d\gamma}{dt}=(k_{1}+k_{2}+k_{3} )n\,,
    \label{eq:dgamma_dt_PN}
\end{equation}
Recall that we have tackled  1PN version of above equation, namely $d \gamma/dt = k_1\,n$
in subsection \ref{sec:orbital_evolution}. This appendix extends such a solution 
while incorporating 2PN and 3PN contributions to the rate of periapsis advance.
The fact that this rate is independent of $\gamma$ allows us to express 
our Eq.~(\ref{eq:dgamma_dt_PN}) as
\begin{equation}
    \frac{d\gamma_j}{dt} = k_j n\,,
\end{equation}
where
\begin{equation}
    \gamma = \sum_{j=1}^{\infty} \gamma_j\,.
\end{equation}

At the 2PN order, we have \cite{KonigsdorfferGopakumar2006}
\begin{equation}
    k_{2}=\left(\frac{GMn}{c^{3}}\right)^{4/3}\frac{\left((51-26\eta)e_{t}^{2}-28\eta+78\right)}{4\left(1-e_{t}^{2}\right)^{2}}\,.
    \label{eq:k2}
\end{equation}
This leads to 
\begin{equation}
\frac{d\gamma_2}{dt} = \frac{1}{4}\left(\frac{GMn}{c^{3}}\right)^{4/3}n^{}\frac{\left((51-26\eta)e_{t}^{2}-28\eta+78\right)}{\left(1-e_{t}^{2}\right)^{2}} \,.  
\end{equation}
Introducing $\tau$ variable with the help of  Eqs.~(\ref{eq:n(e)}), (\ref{eq:P_coeff}) and (\ref{eq:de/dtau})  allow us to write   
\begin{equation}
   \frac{d\gamma_{2}}{d\tau}=-\beta_{2}\frac{\left(1-e_{t}^{2}\right)^{\frac{3}{2}}}{e_{t}^{\frac{42}{19}}}\frac{\left((51-26\eta)e_{t}^{2}-28\eta+78\right)}{\left(304+121e_{t}^{2}\right)^{\frac{3045}{2299}}}\,,
\end{equation}
where 
\begin{equation}
\beta_{2}=\frac{15}{4}\left(\frac{GM_{\text{ch}}n_{0}}{c^{3}}\right)^{-5/3}\left(\frac{GMn_{0}}{c^{3}}\right)^{4/3}\frac{\sqrt{1-e_{t0}^{2}}}{e_{t0}^{6/19}\left(121e_{t0}^{2}+304\right)^{435/2299}}\,.
\end{equation}
We now introduce 
 $\bar{\gamma}_{2}=\Gamma_{20}-\frac{\gamma_{2}}{\beta_{2}}$, where $\Gamma_{20}$ is a constant.
 The above equation then becomes
\begin{equation}
   \frac{d\bar{\gamma}_{2}}{d\tau}=\frac{\left(1-e_{t}^{2}\right)^{\frac{3}{2}}}{e_{t}^{\frac{42}{19}}}\frac{\left((51-26\eta)e_{t}^{2}-28\eta+78\right)}{\left(304+121e_{t}^{2}\right)^{\frac{3045}{2299}}}\,. \label{eq:dgammabar2/dtau}
\end{equation}
We define  ${\gamma}_2$ such that $\gamma_2(\tau_0)=0$.
This allows us to fix $\Gamma_{20}$ to be $\Gamma_{20}=\bar{\gamma}_{2}(\tau_0)$. 
We move on to obtain $ d \bar \gamma_2/d e_t $ by 
dividing Eq.~(\ref{eq:dgammabar2/dtau}) by Eq.~(\ref{eq:de/dtau}), which gives us 
\begin{equation}
    \frac{d\bar{\gamma}_{2}}{de_t}=\frac{e_{t}^{2}(51-26\eta)-28\eta+78}{e_{t}^{13/19}\left(121e_{t}^{2}+304\right)^{\frac{1864}{2299}}}\,.
\end{equation}
This can be integrated to obtain 

\begin{equation}
\bar{\gamma}_{2}(e_t)=\frac{e_t^{6/19}}{336}\left(4\left(121e_t^{2}+304\right)^{\frac{435}{2299}}(51-26\eta)+3\times 2^{\frac{1740}{2299}}19^{\frac{435}{2299}}(2\eta+23)\,_{2}\text{F}_{1}\left(\frac{3}{19},\frac{1864}{2299};\frac{22}{19};\frac{-121e_t^{2}}{304}\right)\right)\,.
\end{equation}
Few comments are in order at this point. 
It should be obvious that we are splitting the GW emission-induced temporal evolution for $\gamma$ in parts. 
This is mainly because we assume that the GW emission is fully prescribed by Eqs.~(\ref{eq:ode_system}). 
And, it explains 
 why we divided $ d \bar \gamma_1/d \tau $  and $ d \bar \gamma_2/d \tau $ equations 
the same equation, namely Eq.~(\ref{eq:de/dtau}) for $d e_t/d\tau$.
In other words, the above split and our division of the resulting 
equations by Eq.~(\ref{eq:de/dtau}) is rather inconsistent if there are higher order contributions to GW emission. 

With the help of these considerations, we move on to write 3PN contributions to $d \gamma/dt$ as 
$d \gamma_3/ dt = k_3 \, n $ where
\begin{align}
   k_{3}&=\left(\frac{GMn}{c^{3}}\right)^{2}\frac{1}{128(1-e_{t}^{2})^{3}}\left(18240-25376\eta+492\pi^{2}\eta+896\eta^{2}+(28128-27840\eta+123\pi^{2}\eta+5120\eta^{2})e_{t}^{2}\vphantom{\sqrt{1-e_{t}^{2}}}\right.\nonumber\\
   &\quad\left.+(2496-1760\eta+1040\eta^{2})e_{t}^{4}+\left(1920-768\eta+(3840-1536\eta)e_{t}^{2}\right)\sqrt{1-e_{t}^{2}}\right)\,.
   \label{eq:k3}
\end{align}
Following the steps that we pursued at the 2PN order lead us to 
\begin{align}
\frac{d\bar{\gamma}_{3}}{de_{t}}&=\frac{1}{e_{t}^{\frac{25}{19}}\left(121e_{t}^{2}+304\right)^{\frac{2734}{2299}}}\left(18240-25376\eta+492\pi^{2}\eta+896\eta^{2}+(28128-27840\eta+123\pi^{2}\eta+5120\eta^{2})e_{t}^{2}\vphantom{\sqrt{1-e_{t}^{2}}}\right.\nonumber\\
&\quad\left.+(2496-1760\eta+1040\eta^{2})e_{t}^{4}+\left(1920-768\eta+(3840-1536\eta)e_{t}^{2}\right)\sqrt{1-e_{t}^{2}}\right)\,,
\end{align}
where we have defined 
$\bar{\gamma}_3=\Gamma_{30}-\frac{\gamma_3}{\beta_3}$, $\Gamma_{30} = \bar{\gamma}_3(\tau_0)$.
Further, the coefficient $\beta_3$ is given by 
\begin{equation}
\beta_{3}=\frac{15}{128}\left(\frac{GM_{\text{ch}}n_{0}}{c^{3}}\right)^{-5/3}\left(\frac{GMn_{0}}{c^{3}}\right)^{2}\frac{e_{t0}^{6/19}\left(121e_{t0}^{2}+304\right){}^{435/2299}}{\sqrt{1-e_{t0}^{2}}}  \;. 
\end{equation}
The equation for $\bar \gamma_3 $ can be solved to get
\begin{align}
\bar{\gamma}_{3}(e_{t})=&-\frac{2957312\sqrt{1-e_{t}^{2}}(2\eta-5)\Appell\left(\frac{-3}{19};\frac{-1}{2},\frac{2734}{2299};\frac{16}{19};e_{t}^{2},\frac{-121e_{t}^{2}}{304}\right)}{e_{t}^{6/19}\left(121e_{t}^{2}+304\right)^{\frac{2734}{2299}}}\nonumber\\&\qquad\times\Bigg[e_{t}^{2}\left(1444\Appell\left(\frac{16}{19};\frac{1}{2},\frac{2734}{2299};\frac{35}{19};e_{t}^{2},\frac{-121e_{t}^{2}}{304}\right)+1367\Appell\left(\frac{16}{19};\frac{-1}{2},\frac{5033}{2299};\frac{35}{19};e_{t}^{2},\frac{-121e_{t}^{2}}{304}\right)\right)\nonumber\\&\qquad\qquad-2432\Appell\left(\frac{-3}{19};\frac{-1}{2},\frac{2734}{2299};\frac{16}{19};e_{t}^{2},\frac{-121e_{t}^{2}}{304}\right)\Bigg]^{-1}\nonumber\\&+\frac{2425920e_{t}^{32/19}\sqrt{1-e_{t}^{2}}(2\eta-5)\Appell\left(\frac{16}{19};\frac{-1}{2},\frac{2734}{2299};\frac{35}{19};e_{t}^{2},\frac{-121e_{t}^{2}}{304}\right)}{\left(121e_{t}^{2}+304\right)^{2734/2299}}\nonumber\\&\qquad\times\Bigg[e_{t}^{2}\left(1444\Appell\left(\frac{35}{19};\frac{1}{2},\frac{2734}{2299};\frac{54}{19};e_{t}^{2},\frac{-121e_{t}^{2}}{304}\right)+1367\Appell\left(\frac{35}{19};\frac{-1}{2},\frac{5033}{2299};\frac{54}{19};e_{t}^{2},\frac{-121e_{t}^{2}}{304}\right)\right)\nonumber\\&\qquad\qquad-5320\Appell\left(\frac{16}{19};\frac{-1}{2},\frac{2734}{2299};\frac{35}{19};e_{t}^{2},\frac{-121e_{t}^{2}}{304}\right)\Bigg]^{-1}\nonumber\\&+\frac{1}{53760\times 2^{1740/2299}19^{435/2299}e_{t}^{6/19}}\nonumber\\&\qquad\times\Bigg[105e_{t}^{2}\left(5120\eta^{2}+3\left(41\pi^{2}-9280\right)\eta+28128\right)\GaussF\left(\frac{16}{19},\frac{2734}{2299};\frac{35}{19};\frac{-121e_{t}^{2}}{304}\right)\nonumber\\&\qquad\qquad-2240\left(224\eta^{2}+\left(123\pi^{2}-6344\right)\eta+4560\right)\GaussF\left(\frac{-3}{19},\frac{2734}{2299};\frac{16}{19};\frac{-121e_{t}^{2}}{304}\right)\nonumber\\&\qquad\qquad+768e_{t}^{4}\left(65\eta^{2}-110\eta+156\right)\GaussF\left(\frac{2734}{2299},\frac{35}{19};\frac{54}{19};\frac{-121e_{t}^{2}}{304}\right)\Bigg] \,.
\label{eq:gamma3bar}
\end{align}

Let us note again that we write  
\begin{equation}
    \gamma(e_{t})=\gamma_{0}-\beta_{1}\left(\bar{\gamma}_{1}(e_{t})-\bar{\gamma}_{1}(e_{t0})\right)-\beta_{2}\left(\bar{\gamma}_{2}(e_{t})-\bar{\gamma}_{2}(e_{t0})\right)-\beta_{3}\left(\bar{\gamma}_{3}(e_{t})-\bar{\gamma}_{3}(e_{t0})\right)\,.
\end{equation}
as we strictly assume that the GW emission is fully characterised by our quadrupolar order equations.

\section{Reactive Evolution of Circular Orbits}
\label{sec:circular_orbits}
This appendix lists the circular limit of GW phasing equations, detailed in Sec.~\ref{sec:orbital_evolution}.  
A careful treatment is required as $\kappa \rightarrow 0$ for circular orbits. 
However, we may obtain $e_t\rightarrow 0$ limit of  Eqs.~(\ref{eq:ode_system}) and it reads
\begin{subequations}
\begin{align}
\frac{dn}{dt}&=\frac{96}{5}\left(\frac{GM_{\text{ch}}n}{c^{3}}\right)^{\frac{5}{3}}n^{2}\,,\label{eq:dn_dt_circ}\\
\frac{de_{t}}{dt}&=0\,,\\
\frac{dl}{dt}&=n\,,\\
\frac{d\gamma}{dt}&=3\left(\frac{GMn}{c^{3}}\right)^{2/3}n +\frac{(78-28\eta)}{4}\left(\frac{GMn}{c^{3}}\right)^{4/3}n+\frac{\left(896\eta^{2}+492\pi^{2}\eta-26144\eta+20160\right)}{128}\left(\frac{GMn}{c^{3}}\right)^{2}n \,,
\end{align}
\end{subequations}
where we have included the 2PN and 3PN contributions to $d\gamma/dt$, using the circular limits of Eqs. (\ref{eq:k2}) and (\ref{eq:k3}).
Since the periapsis is not well-defined for a circular orbit, it is advisable to define the angular variable $\lambda=l+\gamma$ and the sidereal orbital frequency $n_s=(1+k)n$. 
It is straightforward to see that, in terms of $\lambda$ and $n_s$, the orbital evolution can be written as
\begin{subequations}
\begin{align}
\frac{dn_s}{dt}&=\frac{96}{5}\left(\frac{GM_{\text{ch}}n_s}{c^{3}}\right)^{\frac{5}{3}}n_s^{2}\,,\label{eq:dn_dt_circ}\\
\frac{d\lambda}{dt}&=n_s\,,
\end{align}
\end{subequations}
where we have restricted the reactive evolution to the leading order in the PN expansion.

These equations lead to the following analytic expressions for $n_s(t)$ and $\lambda(t)$
\begin{subequations}
\begin{align}
n_{s}(t)&=\frac{n_{s0}}{\left(1-\frac{256}{5}\left(\frac{GM_{\text{ch}}n_{s0}}{c^{3}}\right)^{\frac{5}{3}}n_{s0}(t-t_{0})\right)^{3/8}}\,,\\\lambda(t)&=\lambda_{0}+\frac{1}{32}\left(\frac{GM_{\text{ch}}n_{s0}}{c^{3}}\right)^{-\frac{5}{3}}\left(1-\left(\frac{n_{s0}}{n_{s}}\right)^{5/3}\right)\,,
\end{align}
\end{subequations}
where $n_{s0}$ and  $\lambda_0$ are the values of these variables at some 
initial epoch $t=t_0$.  The orbital eccentricity does not vary in time and its value is chosen to be zero. 
 
The GW emission-induced merger time is obtained by demanding that $n \rightarrow \infty $  and this allows us to write 
\begin{equation}
t_{\text{merg}}=t_{0}+\frac{5}{256n_{0}}\left(\frac{GM_{\text{ch}}n_{0}}{c^{3}}\right)^{-\frac{5}{3}}  \,.  \label{eq:t_merger_circ}
\end{equation}
We have verified that our eccentric version of the merger time,  given by Eq.~(\ref{eq:merger_time_ecc}), reduces to Eq.~(\ref{eq:t_merger_circ}) in the circular limit.

\end{widetext}


\bibliographystyle{apsrev4-1}
\bibliography{GWecc-paper}

\end{document}